%% file: 0_main.tex
\PassOptionsToPackage{table}{xcolor}

\documentclass[letterpaper,twocolumn,10pt]{article}
\usepackage{usenix2019_v3}

\usepackage{tikz}

\usepackage{amsmath}
\usepackage{algorithmic}
\usepackage{graphicx}
\usepackage{textcomp}
\usepackage{xcolor}
\def\BibTeX{{\rm B\kern-.05em{\sc i\kern-.025em b}\kern-.08em
    T\kern-.1667em\lower.7ex\hbox{E}\kern-.125emX}}
    
\usepackage{multirow}
\usepackage{wasysym}
\usepackage{threeparttable}
\usepackage[utf8]{inputenc}
\usepackage[framemethod=TikZ]{mdframed}
\usepackage{tcolorbox}
\usepackage{graphics}
\usepackage{booktabs}
\usepackage{hyperref}
\usepackage{makecell}
\usepackage{textcomp}
\usepackage{enumitem}
\usepackage{listings}
\usepackage{xurl}

\usepackage{pifont}
\newcommand{\cmark}{\ding{51}}%
\newcommand{\xmark}{\ding{55}}%

\usepackage{filecontents}

\begin{document}

\date{}

\title{Is It a Trap? A Large-scale Empirical Study And Comprehensive Assessment of Online Automated Privacy Policy Generators for Mobile Apps}

\author{
{\rm Shidong Pan\thanks{Shidong.Pan@anu.edu.au, College of Engineering, Computing \& Cybernetics, Australian National University (ANU), Australia.}}\\
ANU \& CSIRO's Data61
\and
{\rm Dawen Zhang}\\
CSIRO's Data61 \& ANU
\and
{\rm Mark Staples}\\
CSIRO's Data61
\and
{\rm Zhenchang Xing}\\
CSIRO's Data61 \& ANU
\and
{\rm Jieshan Chen}\\
CSIRO's Data61
\and
{\rm Xiwei Xu}\\
CSIRO's Data61
\and
{\rm Thong Hoang\thanks{Corresponding author. James.Hoang@data61.csiro.au, Software Systems Research Group, CSIRO'Data61, Australia.}}\\
CSIRO's Data61
} 

\maketitle

\begin{abstract}
Privacy regulations protect and promote the privacy of individuals by requiring mobile apps to provide a privacy policy that explains what personal information is collected and how these apps process this information. However, developers often do not have sufficient legal knowledge to create such privacy policies. Online Automated Privacy Policy Generators (APPGs) can create privacy policies, but their quality and other characteristics can vary. In this paper, we conduct the first large-scale empirical study and comprehensive assessment of APPGs for mobile apps. Specifically, we scrutinize 10 APPGs on multiple dimensions.
We further perform the market penetration analysis by collecting 46,472 Android app privacy policies from Google Play, discovering that nearly 20.1\% of privacy policies could be generated by existing APPGs. 
Lastly, we point out that generated policies in our study do not fully comply with GDPR, CCPA, or LGPD.
In summary, app developers must carefully select and use the appropriate APPGs with careful consideration to avoid potential pitfalls.
\end{abstract}

\input{1_Introduction}

\input{3_APPGs}

\input{4_Market}

\input{5_Analysis}

\input{6_Discussion}
\input{7_RelatedWork}
\input{8_Conclusion}

\input{10_Acknowledgement}

\bibliographystyle{plain}
\bibliography{11_Reference}

\input{12_Appendix}

\end{document}

%% file: 1_Introduction.tex
\section{Introduction}
\label{sec_introduction}

Mobile phones and apps are now a ubiquitous part of digital life. There is a large variety and volume of data collected and used by mobile apps, which inevitably brings many privacy issues~\cite{TSE_2018, PETS_2019, liu2021have, USENIX_2022_GEODIFF}. Privacy policies inform users about what, why, and how their personal data are collected and used. These privacy policies have become an important element of responsible technology in mobile app ecosystems.
They also form part of legal agreements for apps and services. Specifically, they are required under regulation in many jurisdictions, such as European General Data Protection Regulation (GDPR)~\cite{GDPR}, California Consumer Privacy Act (CCPA)~\cite{CCPA}, Brazilian Law of General Data Protection (LGPD)~\cite{LGPD}, Australian Privacy Principles (APP)~\cite{APPs}, and Chinese Personal Information Protection Law (PIPL)~\cite{PIPL}.
For example, APP [Art. 1, \textsection{}1.2] imposes an obligation on organisational entities to \textit{``have a clearly expressed and up-to-date APP Privacy Policy about how the entity manages personal information.''}
Privacy policies must be consistent with the data collected and functions provided by service providers, and inadequate policies can create significant business and legal problems.

%
%
\begin{figure*}[t!]
  \centering
  \includegraphics[width=0.95\linewidth]{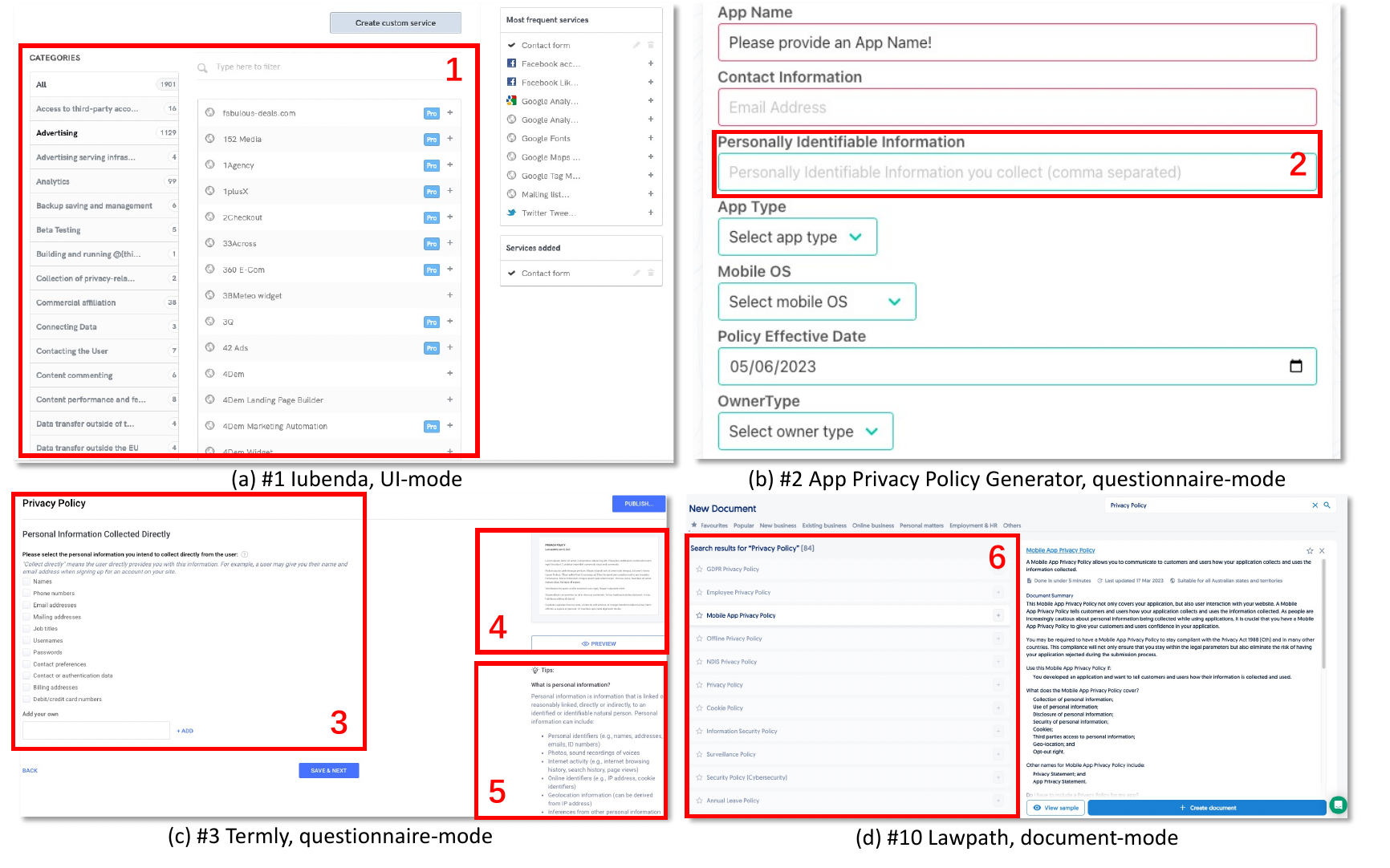}
  \caption{Examples (\#1, \#2, \#3, and \# 10) of APPGs. (1) The integrated UI for users to select privacy practices. (2) The completion question in the questionnaire mode APPGs. (3) The multiple-choice question in the questionnaire mode APPGs. (4) Preview of the generated privacy policy. (5) User instructions. (6) The selection page for document mode APPGs.}
  \label{fig_examples}
\end{figure*}
%
%

Developing privacy policies is a complex process, demanding both the knowledge of app features and corresponding legal requirements.
While larger companies have greater resources and legal expertise to create high-quality privacy policies for their apps, most (citizen) developers do not have legal support and struggle to prepare accurate privacy policies~\cite{balebako2014improving, balebako2014privacy, schaub2015design, li2022understanding}.
To develop privacy policies, developers may copy-paste-modify existing privacy policies, ad-hoc.
As part of this big picture, app development by non-professional developers is growing quickly, supported by trends of using no-code/low-code automated app development tools~\cite{oltrogge2018rise} and pre-trained Large Language Models (LLMs).
To address these needs, Online Automated Privacy Policy Generators (APPGs) can provide more automated solutions and more systematic support for developers to create privacy policies for their apps, rather than through ad-hoc reuse.
Figure~\ref{fig_examples} provides several illustrative examples.
Most APPGs are questionnaire-based tools, which work by asking app developers a series of privacy-related questions about the app, and using those answers to generate privacy policies.
APPGs can create privacy policies, but their quality and other characteristics can vary and are not yet deeply understood.

As we observe in this paper, many apps fail to provide a privacy policy, perfunctorily provide only a low-quality privacy policy, or provide a privacy policy not in local language.
APPGs are becoming an increasingly popular solution used by developers, but developers can be unaware of hidden issues in APPGs~\cite{EASE_2020, zimmeck2021privacyflash}.
Potential design flaws may reflect in the policies, amplify vulnerabilities, violate users' trust assumptions, and ultimately harm end-users.
Thus, a comprehensive and systematic assessment of the capability and limitations of APPGs is necessary.
To better understand the scale of potential problems and their broader impact, it is worthwhile to conduct a market penetration analysis on mainstream APPGs.
Moreover, by scrutinizing the features of popular APPGs, we can glean insights into the demands and preference of both the market and developers.

Numerous previous studies have analysed privacy policies \cite{levy2005improving, ACL_2016, amos2021privacy, nokhbeh2021large, windl2022automating, TSE_2018, PETS_2019, liu2021have, USENIX_2022_GEODIFF, JURIX_2020, EASE_2020} from various perspectives. However, the majority of existing compliance analyses in relation to privacy regulations lack fine-grained scrutiny of specific clauses and requirements. 
APPGs often claim the generated policies are compliant with privacy regulations. 
A more nuanced analysis that hones in on the specific requirements stipulated by privacy regulations is important for a deeper understanding of APPGs' quality.
Other issues relate to mobile apps' dependence on device permissions, for instance, \textit{LOCATION} and \textit{CAMERA}, to function normally.
Generally, we find users are more vigilant when directly providing personal information, but underestimate the impact brought by device permission requests, which when granted by a user allow apps to continuously collect critical personal information without further user approval or consent. 
Therefore, it is significant to examine whether developers correctly display the needed device permissions in the generated privacy policies with APPGs for mobile apps.
\begin{figure*}[t!]
  \centering
  \includegraphics[width=0.8\linewidth]{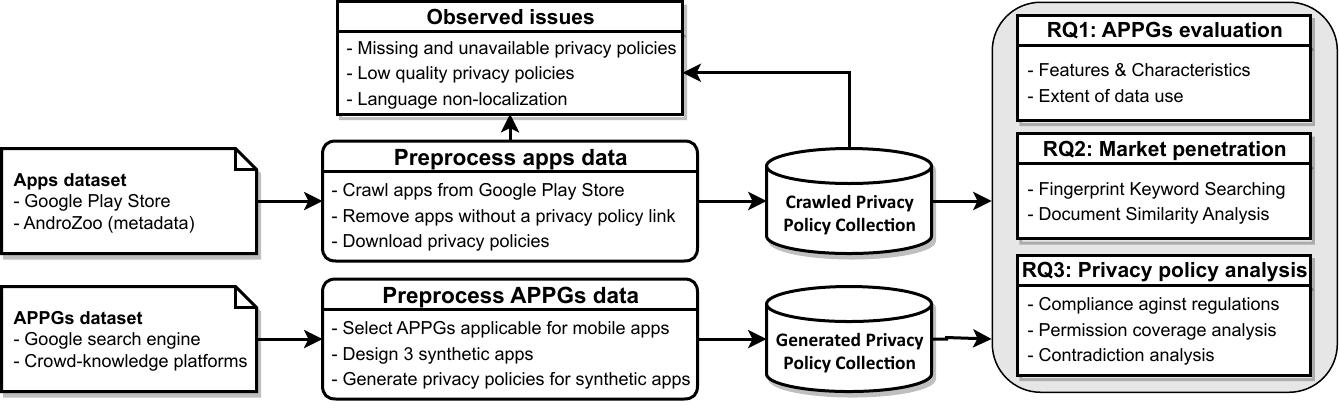}
  \caption{Overview of data flow in the methodology for our empirical study.}
  \label{fig_pipeline}
\end{figure*}
To evaluate current APPGs and their use, we investigate the following research questions: 

\noindent \textbf{RQ1:} What APPGs exist for mobile apps? What are the differences between them? (Section~\ref{sec_APPG})

\noindent \textbf{RQ2:} How many mobile apps' privacy policies could be generated by APPGs? Why are some APPGs more popular? (Section~\ref{sec_market})

\noindent \textbf{RQ3:} To what extent do APPGs comply with privacy regulations in terms of specific requirements, data rights, device permission disclosure, and self-integrity? (Section~\ref{sec_analysis})

An overview of our methodology is shown in Figure~\ref{fig_pipeline}. 
We first collected an initial set of APPGs. 
After removing duplicates and filtering out irrelevant results, we identified 10 APPGs for mobile apps as our research objects.
We conducted an assessment on multiple dimensions, including their features and the recognition of the extent of data use. 
We then conducted a large-scale market penetration analysis.
Based on a random sample of apps from the Google Play app store, we successfully downloaded 46,472 privacy policies (Crawled Privacy Policy Collection) as a large-scale dataset for mobile apps. We found that 15.7\% of apps do not provide a privacy policy link, 22.3\% of them are low-quality privacy policies, and 20.5\% of these privacy policies fail to provide a privacy policy in English, the first language in the target market. 
We then generated 30 privacy policies (Generated Privacy Policy Collection) using the 10 APPGs for three synthetic apps and conducted a market penetration analysis. 
Our results show 20.1\% app developers favorably use APPGs to generate their privacy policies, and \#2 \textit{App Privacy Policy Generator} is the most popular one, boasting a 72.7\% adoption rate. 
In an effort to more accurately assess compliance with privacy regulations, we scrutinized the generated privacy policies.
Our findings revealed a substantial level of noncompliance with privacy laws and a frequent under-claiming of data rights and highly concerning privacy practices, especially the most popular \#2.
In addition, our results suggest dangerous permissions are commonly missing in the generated policies of APPGs, and inappropriate APPGs selection will hinder developers' capacity to include essential device permissions.
Furthermore, we found that APPGs might introduce more privacy policy contradiction issues, undermining the self-integrity.

\textbf{Contributions.}
To our knowledge, this is the first large-scale, exhaustive empirical study of automated privacy policy generators for mobile apps. Our observations, findings, and insights will benefit stakeholders beyond just app developers, including APPG providers and privacy regulators from the APPG design perspective and legal perspective.
We make the following contributions:
\begin{itemize} [leftmargin=*]
    \item We perform a systematic empirical analysis of APPGs, covering various aspects, such as features, characterizations, and levels of recognition of data use.

    \item We conduct a systematic analysis of the privacy policies of 99,149 apps in the Google Play Store. Specifically, our dataset includes 46,472 privacy policies extracted from these apps and available at~\cite{our_repo}. 

    \item We discuss implications for APPG users, APPG providers, and privacy regulators to improve the APPG ecosystem.   
\end{itemize}

%% file: 3_APPGs.tex
\section{Online Automated Privacy Policy Generators for Mobile Apps (\textbf{RQ1})}
\label{sec_APPG}
This section addresses RQ1, describing 
what APPGs exist for mobile apps and what differences there are between them. We manually collected and identified 10 publicly available online APPGs for mobile apps. We analysed various characteristics and specifically the range of possible data uses in apps.

\subsection{Collecting Online APPGs for Mobile Apps}
\label{sec_APPG_collect}

To better understand the current status of APPGs, we identified popular APPGs for mobile apps by using the Google search engine and exploring crowd-knowledge platforms.
In detail, we employed the Google search engine to broadly search APPGs for mobile apps. 
We acted as a hypothetical citizen developer who is trying to generate a privacy policy for their app. 
We created an initial set of search terms, including ``\textit{Apps Privacy Policy Generators}'', ``\textit{Apps Privacy Policy Generators Online}'', ``\textit{Privacy Policy Automated Generation}'', ``\textit{Google Play Store Privacy Policy Generator}'', ``\textit{GDPR Policy Generators}''. 
For each search keyword, we evaluated all entries on the first result page. In addition to direct results from Google search, we also identified APPGs by using our Google search terms again in app developer crowd-knowledge platforms including Stack Overflow~\cite{stackoverflow}, GeeksforGeeks~\cite{geeksforgeeks}, and GitHub~\cite{github}. 
To evaluate search results, we manually identified and read highly related threads, and collected the mentioned APPGs.
After collecting an initial set of APPGs and removing duplicates, we filtered out irrelevant APPGs based on the following two criteria: 1) APPGs that only support privacy policy generation for websites; and 2) APPGs that do not actually provide their proposed functions.
For example, while some APPGs claim that they are able to generate privacy policies for mobile apps, they do not actually provide such a service.

Table~\ref{tab_overview} presents 10 identified APPGs in our dataset.
``Freeware'' represents whether a generator needs monetary investment to use. 
``Free'' means the APPG provides a free version and ``Paid'' means the APPG provides a premium version that unlocks more features. 
``Registration'' indicates whether a user needs to register an account to use the APPG. 
All APPGs we collected are based on question answering and template completion (`boilerplate'), however, they have various different ``Modes'' to handle information interactions with users as we demonstrated in Figure~\ref{fig_examples}. 
``User interface''(UI) mode means that users need to select the relative data practices through an integrated UI as à la carte manner, then answer a few other questions about the app's and developers' basic information. 
``Questionnaire'' mode only requires users to complete a long questionnaire that covers all potential data practices and a privacy policy will be generated depending on the answers.
This is the most common and popular type of APPGs (8/10).
``Document'' mode is common in legal consultation websites. Users need to purchase access to a privacy policy document, then answer a series of questions to complete the template. 
Notably, all analyses of APPGs in this paper include features in both free and paid versions. We conducted all analyses of APPGs in May 2022.

\input{Tables/Overview}
%
%

\subsection{Characterization}
\label{sec_characterization}

A comprehensive assessment is conducted on identified APPGs along 10 dimensions. The first five (1-5) target practicability, user-friendliness, and convenience from potential APPG users' perspectives. The next three (6-8) cover legal compliance with privacy regulations, according to the tools' claims. The last two (9, 10) focus on the understandability of privacy policies. 
The results are shown in Table~\ref{tab_characterization}. We define these 10 dimensions as follows:

\input{Tables/Characterization}

\textbf{\texttt{Extent of Data Recognised (1)}}. Mobile apps normally require data from direct user input or device sensors. 
Some apps may also provide data to third-party services, for revenue or to enhance the user experience. 
Privacy regulations such as GDPR [Art. 14(1)(d)] and CCPA [\textsection{}1798.130(a)(5)(B), 1798.110(c), Regs \textsection{}999.308(c)(1)(d)))], can require that app developers include information about all collected or shared data within their privacy policy. 
However, some APPGs only recognise a limited range of data practices for inclusion in privacy policies. 
As failing to include all data practices could be regarded as a regulatory violation in some jurisdictions, this is an important characteristic and is discussed in more detail in the next section (Section~\ref{sec_diversification}).
Tentatively, we use (\CIRCLE) to denote that the APPG recognises a wide range of data use, (\LEFTcircle) for a smaller range of data use, and (\Circle) for a few kinds of data use. 

\textbf{\texttt{Customizability (2)}}. Customizability addresses two concerns: can users add privacy practices that were not originally included in the APPG pipeline, and do users have the freedom to add customized clauses in the generated privacy policy document? 
The former can help developers to provide a more specific and accurate privacy policy; the latter can more easily accommodate developers who want to address additional concerns about users' privacy. 
APPGs are denoted as (\CIRCLE), (\LEFTcircle), and (\Circle), as indications that they are positive for both, one, and zero of these questions, respectively.

\textbf{\texttt{User Instruction (3)}}. The main target audience of APPGs is expected to be developers with little or no legal knowledge about privacy and data protection.  Therefore it is crucial to provide sufficient user instructions to help correctly utilize the APPG.
User instructions can be generally categorised into three types: providing further explanation of obscure questions (an example is shown in Figure~\ref{fig_examples}), elaborating terms by listing examples, and supporting users with an interactive help center. 
APPGs are scored as (\Circle) if they only provide one or nil occurrence of the above user instructions, (\LEFTcircle) for two to five occurrences, and (\CIRCLE) for more than five occurrences. Especially, APPG \#1 \textit{Iubenda} also provides an introduction video and comprehensive documentation.

\textbf{\texttt{Complexity (4)}}. This dimension indicates the general learning cost for APPG users, and higher complexity reflects a higher cost.
For UI-mode APPGs, users have to spend more time getting familiar with the UI and learning operation procedures, therefore, we manually set them as (\CIRCLE).
For questionnaire-mode APPGs, we found that the complexity is related to the number of questions that users are asked in the generation process.
Based on the ``Statistic summary'' in Table~\ref{tab_diversification}, if the sum of maximum multiple-choice questions and maximum completion is less than 10, the APPG is marked as (\Circle), if it is between 10 to 50, the APPG is marked as (\LEFTcircle); and if it is greater than 50, the APPG is marked as (\CIRCLE).
For document-mode APPGs, users only need to select a document from the library and fill up several questions about their basic information, thus, we manually set them as (\Circle).

\textbf{\texttt{Publishing Support (5)}}. This dimension reflects how convenient it is to deploy the generated privacy policy.
Every APPG provides at least one of the following publishing options: 1) a permanently hosted website link containing the privacy policy; and 2) the generated privacy policy context in editable HTML format.
In section~\ref{sec_observation_missing}, we presented that a large amount of privacy policy links in the market lead to inaccessible websites, and we believe that a permanently hosted website link provided by APPGs can mitigate this problem to some extent.
(\CIRCLE) indicates the APPG provides both options and (\LEFTcircle) means APPGs only provide option 2.

\textbf{\texttt{GDPR (6), CCPA (7), and LGPD (8)}}. These three dimensions are binary, simply indicating whether the APPG claims that they provide compliance with the corresponding privacy regulations. (\CIRCLE) denotes support and (\Circle) denotes non-support. No APPG claims compliance with other privacy regulations such as APP~\cite{APPs} and PIPL~\cite{PIPL}.

\textbf{\texttt{Multilingual Support (9)}}. In Section \ref{sec_observation_language}, we discuss how current privacy policies suffer from the language localization problems. 
APPGs that provide multi-language support should benefit both app developers and app users. 
Additionally, in CCPA [Regs §999.308(a)(2)(d)], it stipulates that online notices should follow generally recognized industry standards, such as the W3C Web Content Accessibility Guidelines~\cite{WCAG}.
We use (\CIRCLE) to denote that APPG can generate privacy policies in more than one language, otherwise (\Circle).

\textbf{\texttt{Readability (10)}}. 
While a long privacy policy can provide more comprehensive and detailed descriptions of data practices in apps, adding unnecessary information to a privacy policy may lead to information overload~\cite{bawden2009dark, waldman2020cognitive}.
Although APPGs can generate more concise privacy policies that are more specifically tailored to the data and data practices relevant to each app, this does not necessarily mean that those privacy policies are more readable.
Some regulations require mobile app developers to use clear and understandable language in privacy statements~\cite{voss2014looking, krumay2020readability}. 
Also, W3C Web Content Accessibility Guidelines [\textsection{}3.1] requires that ``make text content readable and understandable''.
However, those high-level principles do not give details for this requirement.
Thus, we adopt the Flesch Reading-Ease Test~\cite{kincaid1975derivation} to evaluate the readability of generated privacy policy context.
A higher score indicates the content is easier to read and understand, the score scales from 0 to 100.
The average readability score of the privacy policies of 12 leading mobile apps~\cite{leadingapps}, who have over one billion installs, is 46.
We use (\CIRCLE) when the readability score is greater than or equal to 46, (\LEFTcircle) for 40 to 46, and (\Circle) for less than 40. Specific readability scores are listed in Appendix Table~\ref{tab_readability}.

Since the assessment of apps in these dimensions involved some subjective judgments, the first two authors assessed the 10 APPGs individually. For any disagreement at the table, they discussed and agreed on the same answer, and if the disagreement persisted, a third author joined the discussion to facilitate a resolution.
As shown in Table~\ref{tab_characterization}, the level of Complexity is positively associated with the level of Extent of Data Use,
Customizability, and User Instruction. 
Most of the APPGs (8/10) exhibit complete publishing support.
As for compliance with privacy regulations, more than half the APPGs (6/10) claim that their privacy policies conform to the GDPR and CCPA, and only \#1 \textit{Iubenda} claims compliance with LGPD. 
Only \#1 \textit{Iubenda} supports generating privacy policies in languages other than English.
All APPGs' readability scores are between 30 and 50, which indicates that readers should have at least a college-level education background to easily read the policies~\cite{kincaid1975derivation}.
\#2 \textit{App Privacy Policy Generator} has the highest readability score and is the most frequently used APPG on the current market, as discussed in Section~\ref{sec_analysis_market}. 

\input{Tables/APPG_details}

\subsection{Extent of Data Use Recognised by APPGs}
\label{sec_diversification}

In this section, we further investigate the extent of data, data use, and data practises recognised by APPGs to support the generation of privacy policies.
We identified four major aspects: the app’s basic information, users’ personal information, device permissions, and third-party services.
We carefully gathered all the possible questions and potential options that APPG users might face.
If any question or option explicitly pertains to a specific data type, permission, or third-party service, then it is classified as ``recognised'' (\CIRCLE), otherwise ``absent'' (\Circle).
The specific item enumeration process is available at~\cite{our_repo}.



\textbf{\texttt{App’s basic information}}. \hspace{-5pt} Apart from user names, providing at least one communication channel to app users for any potential inquiries is commonly required by privacy protection and regulation laws, and all the APPGs support this with different levels of granularity. 

\textbf{\texttt{Users personal information}}. \hspace{-5pt} We manually separate this into two sub-categories: general personal information, and sensitive personal information.
Most APPGs recognise general personal information, and only some APPGs (3/10) explicitly identify sensitive information.

\textbf{\texttt{Device permissions}}. According to the Android developer's guide~\cite{android_developer_guide}, there are nine dangerous permission groups, namely \textit{CALENDAR}, \textit{CAMERA}, \textit{CONTACTS}, \textit{LOCATION}, \textit{MICROPHONE}, \textit{PHONE}, \textit{SENSORS}, \textit{SMS}, and \textit{STORAGE}, on mobile phones. We expect the collection and use of data related to these permissions to be clearly declared in a privacy policy, therefore, they are set as row indices. 
We found that more than half of the APPGs (6/10) recognise \textit{CAMERA}, \textit{CONTACTS}, and \textit{LOCATION}, and only two APPGs identify all permissions.
Failing to correctly state sensor permissions in the privacy policy could lead to serious privacy issues, so we further investigated whether users can correctly utilise the APPG to state all claimed sensor permissions in Section~\ref{sec_analysis_permissions}.

\textbf{\texttt{Third-party services}}. Third-party services have become a significant part of mobile apps. Taking into account the diversity of service providers and their popularity, we selected five third-party services as row indices, as most APPGs either fully or partially cover them.
Overall, APPGs show decent coverage among these third-party services.

\textbf{\texttt{Statistic summary}}. We counted the completion questions and multi-choice questions (examples are illustrated in Figure~\ref{fig_examples}) that users would encounter for the questionnaire-mode APPGs.
Although questions may vary dramatically, we take it as a quasi-indicator to approximately reflect users' learning and time cost to use the APPGs.

As it involves intensive manual work, to avoid the effect caused by potential human error, we employ the same strategy as introduced in the previous section.
Cohen's Kappa~\cite{cohen1960coefficient} $\kappa = 0.92$ for the initial manual labelling, which is an almost perfect level of agreement.
As shown in Table~\ref{tab_diversification}, \#1, \#3, and \#8 are the best in terms of the recognised data use, they almost cover every data index listed in the table.
\#2, \#4, \#7, and \#9 decently cover users’ general personal information and third-party services, but only a few or null users’ sensitive personal information and device permissions.
\#5, \#6, and \#10 merely cover some app's basic information and failed to support the rest of the categories. 
Above all, users need to carefully select appropriate APPGs, otherwise, the unsupported data use will be missing in the generated privacy policy.

\begin{tcolorbox}

     \noindent \textbf{Finding 1:} APPGs are a handy solution for developers to draft privacy policies for their apps. They have various features, characterizations, and levels of recognition of data use.
\end{tcolorbox}

%% file: Tables/Overview.tex
\begin{table}[tbp]
\centering
\caption{APPGs for mobile apps identified in this study.}
\label{tab_overview}
\resizebox{0.9\linewidth}{!}{%
\begin{tabular}{rlrrr} 
\toprule

\textbf{\#} & \textbf{Name}                                                          & \textbf{Freeware} & \textbf{Registration} & \textbf{Mode}   \\ 
\midrule
1           & Iubenda~\cite{Iubenda}                                                                & Free\&Paid        & Required              & User Interface  \\
\midrule
2           & \begin{tabular}[c]{@{}l@{}}App Privacy Policy \\Generator~\cite{AppPrivacyPolicyGenerator}~\end{tabular} & Free only         & Not required          & Questionnaire   \\
\midrule
3           & Termly~\cite{termly}                                                              & Free\&Paid        & Required              & Questionnaire   \\
\midrule
4           & Privacy Policies~\cite{privacypolicies}                                                      & Free\&Paid        & Required              & Questionnaire   \\
\midrule
5           & App Privacy Policy~\cite{appprivacypolicy}                                                     & Free only         & Not required          & Questionnaire   \\
\midrule
6           & \begin{tabular}[c]{@{}l@{}}Privacy Policy\\ Online\cite{privacypolicyonline}~\end{tabular}                                                   & Free only         & Not required          & Questionnaire   \\
\midrule
7           & Terms Feed~\cite{termsfeed}                                                             & Free\&Paid        & Required              & Questionnaire   \\
\midrule
8           & Website Policies~\cite{websitepolicies}                                                       & Free\&Paid        & Required              & Questionnaire   \\
\midrule
9           & Free Privacy Policy~\cite{freeprivacypolicy}                                                    & Free\&Paid        & Required              & Questionnaire   \\
\midrule
10          & Lawpath~\cite{lawpath}                                                                & Paid only         & Required              & Document        \\
\bottomrule
\end{tabular}
}
\end{table}

%% file: Tables/Characterization.tex
\begin{table*}
  \caption{The characterization of 10 APPGs on 10 different dimensions.}
  \label{tab_characterization}
  \centering
  \resizebox{1.0\textwidth}{!}{%
  \begin{tabular}{l|ccccc|ccc|cc}
    \toprule
    \textbf{\# and APPG Name} & 
    \rotatebox[origin=c]{45}{\textbf{Extent of Data Use}}&
    \rotatebox[origin=c]{45}{\textbf{Customizability}}& 
    \rotatebox[origin=c]{45}{\textbf{User Instruction}}& 
    \rotatebox[origin=c]{45}{\textbf{Complexity}}& 
    \rotatebox[origin=c]{45}{\textbf{Publishing Support}}& 
    \rotatebox[origin=c]{45}{\textbf{GDPR}}&
    \rotatebox[origin=c]{45}{\textbf{CCPA}}&
    \rotatebox[origin=c]{45}{\textbf{LGPD}}&
    \rotatebox[origin=c]{45}{\textbf{Multilingual Support}}&
    \rotatebox[origin=c]{45}{\textbf{Readability}}\\
    \midrule
    \rowcolor{lightgray!85}
    \#1 Iubenda &\CIRCLE &\LEFTcircle & \CIRCLE &\CIRCLE & \CIRCLE &\CIRCLE &\CIRCLE &\CIRCLE &\CIRCLE &\Circle\\
    
    \rowcolor{lightgray!15}
    \#2 App Privacy Policy Generator &\LEFTcircle &\Circle &\Circle &\Circle &\LEFTcircle &\Circle &\Circle &\Circle &\Circle &\CIRCLE\\
    
    \rowcolor{lightgray!85}
    \#3 Termly &\CIRCLE &\LEFTcircle &\CIRCLE &\CIRCLE &\CIRCLE &\CIRCLE &\CIRCLE & \Circle &\Circle &\CIRCLE\\
    
    \rowcolor{lightgray!15}
    \#4 Privacy Policies &\LEFTcircle &\Circle  & \Circle &\LEFTcircle &\CIRCLE  &\CIRCLE &\CIRCLE & \Circle &\Circle &\Circle\\
    
    \rowcolor{lightgray!85}
    \#5 App Privacy Policy &\Circle &\Circle &\Circle &\Circle &\CIRCLE &\Circle &\Circle &\Circle &\Circle &\LEFTcircle\\

    \rowcolor{lightgray!15}
    \#6 Privacy Policy Online  &\Circle &\Circle & \Circle &\Circle &\CIRCLE &\Circle &\Circle &\Circle &\Circle &\LEFTcircle\\
    
    \rowcolor{lightgray!85}
    \#7 Terms Feed  &\LEFTcircle &\LEFTcircle & \Circle &\LEFTcircle &\CIRCLE &\CIRCLE &\CIRCLE & \Circle &\Circle &\Circle\\
    
    \rowcolor{lightgray!15}
    \#8 Website Policies &\CIRCLE &\CIRCLE &\LEFTcircle &\CIRCLE  &\CIRCLE &\CIRCLE &\CIRCLE & \Circle &\Circle &\Circle\\
    
    \rowcolor{lightgray!85}
    \#9 Free Privacy Policy & \LEFTcircle  &\Circle &\Circle &\LEFTcircle &\CIRCLE &\CIRCLE &\CIRCLE & \Circle &\Circle &\Circle\\

    \rowcolor{lightgray!15}
    \#10 Lawpath &\Circle &\Circle &\CIRCLE &\Circle &\LEFTcircle & \Circle & \Circle & \Circle &\Circle &\LEFTcircle\\
    
  \bottomrule
\end{tabular}%
}
\begin{tablenotes}
  \footnotesize
  \centering
  \item \Circle : low level / does not support;
  \LEFTcircle : intermediate level / partially support;
  \CIRCLE : high level / fully support
\end{tablenotes}
\end{table*}

%% file: Tables/APPG_details.tex
\begin{table*}[t!]
  \caption{The breakdown of APPG's recognition of data use. The first row of each section is the number of APPGs as per Table \ref{tab_overview}. ``Multiple-choice questions'' refers to whether developers select options from listed items, and ``Completion questions'' means developers fill in blanks with plain text. (Selecting a date is also regarded as a completion question.) Since some questions sometimes unlock later questions, we use ``Minimum'' and ``Maximum'' to reflect the lower and upper bounds. *If date of birth is included in the APPG, then we consider that Age group can be inferred.}
  \label{tab_diversification}
  \centering
  \resizebox{0.85\textwidth}{!}{%
  \begin{tabular}{l|ccccc | ccccc}
    \toprule
    \rowcolor{lightgray!85}
    \textbf{App's basic information}& 1 & 2& 3 &4&5& 6 & 7& 8 &9&10\\
    \midrule
    \rowcolor{lightgray!15}
    Name of the app & \CIRCLE & \CIRCLE& \CIRCLE&\CIRCLE&\CIRCLE&\CIRCLE&\CIRCLE&\CIRCLE& \CIRCLE& \CIRCLE\\
    
    \rowcolor{lightgray!15}
    Name of the developer &\CIRCLE &\CIRCLE &\CIRCLE&\CIRCLE &\Circle &\Circle &\CIRCLE &\CIRCLE &\CIRCLE &\CIRCLE\\
    
    \rowcolor{lightgray!15}
    State/Country of the developer& \Circle &\Circle &\CIRCLE&\CIRCLE &\CIRCLE &\CIRCLE & \CIRCLE &\Circle &\CIRCLE &\CIRCLE\\
    
    \rowcolor{lightgray!15}
    Physical address of the developer& \CIRCLE &\CIRCLE &\CIRCLE&\CIRCLE &\Circle &\Circle &\CIRCLE &\CIRCLE &\CIRCLE &\CIRCLE  \\
    
    \rowcolor{lightgray!15}
    Email address of the developer & \CIRCLE &\CIRCLE &\CIRCLE &\CIRCLE&\CIRCLE &\CIRCLE &\CIRCLE &\CIRCLE &\CIRCLE &\CIRCLE\\
    
    \rowcolor{lightgray!15}
    Phone number of the developer &\Circle &\CIRCLE &\CIRCLE &\CIRCLE &\Circle &\Circle &\CIRCLE &\Circle &\CIRCLE &\Circle \\
    
    \rowcolor{lightgray!15}
    Customizable policy effective date & \CIRCLE &\CIRCLE &\CIRCLE &\Circle &\Circle &\Circle &\Circle &\Circle &\Circle &\CIRCLE\\
    \midrule
    
    \rowcolor{lightgray!85}
    \textbf{Users' personal information}& 1 & 2& 3 &4&5& 6 & 7& 8 &9&10\\
    \midrule
    \rowcolor{lightgray!50}
    \textbf{Users' general personal information} & & & && && & & &\\
    \rowcolor{lightgray!15}
    Name &\CIRCLE &\CIRCLE &\CIRCLE &\CIRCLE &\Circle &\Circle &\CIRCLE &\CIRCLE &\CIRCLE &\Circle\\
    \rowcolor{lightgray!15}
    Age group* (adulthood or underage) &\CIRCLE &\Circle &\CIRCLE &\Circle &\Circle &\Circle &\Circle &\CIRCLE &\Circle &\Circle \\
    \rowcolor{lightgray!15}
    Phone numbers & \CIRCLE &\CIRCLE &\CIRCLE &\CIRCLE &\Circle &\Circle &\CIRCLE &\CIRCLE &\CIRCLE &\Circle\\
    \rowcolor{lightgray!15}
    Email address & \CIRCLE&\CIRCLE &\CIRCLE &\CIRCLE &\Circle &\Circle &\CIRCLE &\CIRCLE &\CIRCLE &\Circle\\
    \rowcolor{lightgray!15}
    Residential addresses & \CIRCLE &\CIRCLE &\CIRCLE &\CIRCLE &\Circle &\Circle &\CIRCLE &\Circle &\CIRCLE &\Circle\\

    \rowcolor{lightgray!50}
    \textbf{Users' sensitive personal information} & & & && && & & &\\
    \rowcolor{lightgray!15}
    Health data & \CIRCLE &\Circle &\CIRCLE &\Circle &\Circle &\Circle & \Circle&\CIRCLE &\Circle &\Circle\\
    \rowcolor{lightgray!15}
    Biometric data &\CIRCLE &\Circle &\CIRCLE &\Circle &\Circle &\Circle &\Circle &\CIRCLE &\Circle &\Circle\\
    \rowcolor{lightgray!15}
    Gender data &\CIRCLE &\Circle &\CIRCLE &\Circle &\Circle &\Circle & \Circle &\Circle &\Circle &\Circle\\
    \rowcolor{lightgray!15}
    Information revealing race or ethnic origin &\Circle &\Circle &\CIRCLE &\Circle &\Circle &\Circle &\Circle &\CIRCLE &\Circle &\Circle\\
    \rowcolor{lightgray!15}
    Government identifiers (e.g., medical card number) &\CIRCLE &\Circle &\CIRCLE &\Circle &\Circle &\Circle & \Circle &\CIRCLE &\Circle &\Circle\\
    \midrule
    
    \rowcolor{lightgray!85}
    \textbf{Device permissions}& 1 & 2& 3 &4&5& 6 & 7& 8 &9&10\\
    \midrule
    \rowcolor{lightgray!15}
    Calendar permission &\CIRCLE &\Circle &\CIRCLE &\Circle &\Circle &\Circle & \Circle&\Circle  &\Circle &\Circle \\
    \rowcolor{lightgray!15}
    Camera permission &\CIRCLE &\Circle &\CIRCLE &\CIRCLE &\Circle &\Circle & \CIRCLE&\Circle  &\CIRCLE &\Circle \\
    \rowcolor{lightgray!15}
    Contacts permission &\CIRCLE &\Circle &\CIRCLE &\CIRCLE &\Circle &\Circle & \CIRCLE &\CIRCLE &\CIRCLE &\Circle \\
    \rowcolor{lightgray!15}
    Location permission &\CIRCLE &\Circle &\CIRCLE &\CIRCLE &\Circle &\Circle &\CIRCLE &\CIRCLE &\CIRCLE &\Circle \\
    \rowcolor{lightgray!15}
    Microphone permission &\CIRCLE &\Circle &\CIRCLE &\Circle &\Circle &\Circle & \Circle&\Circle &\Circle &\Circle \\
    \rowcolor{lightgray!15}
    Phone permission &\CIRCLE &\Circle &\CIRCLE &\Circle &\Circle &\Circle &\Circle  &\Circle &\Circle &\Circle \\
    \rowcolor{lightgray!15}
    Sensor permission &\CIRCLE&\Circle &\CIRCLE& \Circle &\Circle &\Circle & \Circle &\Circle &\Circle &\Circle \\
    \rowcolor{lightgray!15}
    SMS permission &\CIRCLE&\Circle &\CIRCLE& \Circle &\Circle &\Circle & \Circle &\Circle &\Circle &\Circle \\
    \rowcolor{lightgray!15}
    Storage permission &\CIRCLE &\Circle &\CIRCLE &\Circle &\Circle &\Circle &\Circle &\Circle &\Circle &\Circle \\
    \midrule

    \rowcolor{lightgray!85}
    \textbf{Selected third-party services}& 1 & 2& 3 &4&5& 6 & 7& 8 &9&10\\
    \midrule
    \rowcolor{lightgray!15}
    Facebook account access &\CIRCLE &\CIRCLE &\CIRCLE &\Circle &\Circle &\Circle &\Circle &\CIRCLE &\Circle &\Circle \\
    \rowcolor{lightgray!15}
    Twitter account access &\CIRCLE&\Circle &\CIRCLE &\Circle &\Circle &\Circle &\Circle &\CIRCLE &\Circle &\Circle \\
    \rowcolor{lightgray!15}
    Google Analytics for Firebase&\CIRCLE&\CIRCLE &\CIRCLE &\CIRCLE &\Circle &\Circle & \CIRCLE &\CIRCLE &\CIRCLE &\Circle \\
    \rowcolor{lightgray!15}
    Flurry Analytics&\CIRCLE&\CIRCLE &\CIRCLE &\CIRCLE &\Circle &\CIRCLE &\CIRCLE &\CIRCLE &\CIRCLE &\Circle \\
    \rowcolor{lightgray!15}
    AdMob&\CIRCLE&\CIRCLE &\CIRCLE &\CIRCLE &\Circle &\CIRCLE  &\CIRCLE &\CIRCLE &\CIRCLE &\Circle \\
    \midrule

    

    \rowcolor{lightgray!85}
    \textbf{Statistic summary}& 1 & 2& 3 &4&5& 6 & 7& 8 &9&10\\
    \midrule
    
    \rowcolor{lightgray!15}
    Minimum multiple-choice questions & / &4 &58 &7 &2 &5 &7 &38 &7 & / \\

    \rowcolor{lightgray!15}
    Maximum multiple-choice questions & /&5 &79 &26 &2 &5 &26 &63 &26 &/  \\

    \rowcolor{lightgray!15}
    Minimum completion questions &/ &4 &11 &2 &2 &3 &2 &2 &2 &/  \\

    \rowcolor{lightgray!15}
    Maximum completion questions &/ &4 &76 &22 &2 &3 &22 &13 &22 & /\\

  \bottomrule
\end{tabular}
}%
\end{table*}

%% file: 4_Market.tex
\section{Market Penetration of APPGs (\textbf{RQ2})}
\label{sec_market}

The previous section carefully examined various characteristics of APPGs. In this section we consider the questions: How large is the APPG market? And are the seemingly more functional APPGs actually more popular in the market?
\subsection {Status Quo of Mobile Apps' Privacy Policy}
\label{sec_overview}

Privacy policies play an essential role in mobile app ecosystems.
An app without a privacy policy is not necessarily malicious, but developers are required to provide privacy policies due to regulations in some jurisdictions. These policies are also increasingly demanded by mobile app users and marketplaces.
According to the Apple App Store Developers page \cite{appledeveloper}, a privacy policy is required when submitting new apps or app updates:
\textit{``By adding the following links on your product page, you can help users easily access your app’s privacy policy and manage their data in your app. \textbf{Privacy Policy (Required):} The URL to your publicly accessible privacy policy.''}
Additionally, in Google Play Console Help Center~\cite{googledeveloper}, developers \textit{``...must include a link to your privacy policy''} to prepare the app for review.
Although platform principles regulate the existence and quality of privacy policies, we still discovered some critical problems in the mobile apps market.
Other forms of privacy notices, such as privacy nutrition labels (PNLs)~\cite{li2022understanding, zhang2022usable, zhangprivacy, pan2023toward}, focus more on providing succinct information to end users, potentially sacrificing information relevant or required by specific regulations or marketplace requirements. 

\subsection{Crawled Privacy Policy Dataset}
\label{sec_overview_dataset}
To better understand the status and quality of existing privacy policies, we need a large-scale privacy policy dataset.
However, existing datasets suffer from various problems. 

\textbf{The differences between datasets for previous research and our research goals.} 
Existing datasets are either not focused on mobile apps, biased, small, or outdated. First, some prior research focuses on privacy policies across all platforms, including websites and mobile apps \cite{amos2021privacy, EASE_2020, CMU_2017}, while we only target mobile apps.
Since they did not include metadata (e.g., platforms) of the collected policies, we cannot filter them out from
other platforms.
Second, some research only focuses on apps of certain groups, and their collected datasets are not representative enough for apps in the market.
For example, one study~\cite{USENIX_2022_GEODIFF} manually collected 5,684 ``globally popular apps.'' 
This selection criterion leads to two critical problems: 
a) The top overall popular apps appear in a limited set of categories, such as \textit{Social} and \textit{Finance}, which can drive out apps from less popular categories such as \textit{Parenting} or \textit{Libraries \& Demo}. 
b) These globally popular apps do not distribute evenly across various app categories, which may cause potential bias and unrepresentative findings.
Third, some existing datasets are small and can contain potential bias.
For example, APP-350~\cite{PETS_2019} only analysed 350 privacy policies, and another study~\cite{liu2021have} conducted fine-grained analysis on only 304 privacy policies.
Finally, due to the rapid iteration of apps, apps' privacy policies, and APPGs, some prior datasets can quickly become outdated.
In summary, to make our findings comprehensive, generalizable, and up-to-date, we needed to collect a current large-scale dataset that can assemble apps of different popularity and from various app categories.

\textbf{Google Play Store.} 
Android mobile apps are often more transparent and friendly to academic research compared to iOS mobile apps~\cite{habchi2017code, kollnig2021iphones}. 
There are many app markets for the Android platform, but Google Play is the largest and the most accessible app market, with over two million apps, according to AppBrain~\cite{appbrain}. Thus, we only focus on it in this study.
Some Android apps are specially designed for Android smartwatches instead of mobile phones; therefore, we intentionally excluded apps belonging to categories named \textit{Watch apps} and \textit{Watch faces}.

\textbf{Our dataset.}
We collected a new large-scale privacy policy dataset from existing apps on the Google Play Store.
We also collected app metadata from AndroZoo~\cite{Allix:2016:ACM:2901739.2903508}, which is a large-scale and growing Android app collection extracted from multiple sources, including the Google Play app store.
We removed duplicates and apps from other app markets and then randomly sampled 268,500 apps as our initial app dataset, which is around 10\% of the whole app population.
From this initial sample, we observed that some apps are either invalid or not usable.
These were mainly dummy apps or student apps with a package size normally less than 10 KB, so we removed invalid apps like these.
We further excluded unavailable apps that returned error messages on their Google Play page. 
Apps can be unavailable for many reasons including geographic differences \cite{USENIX_2022_GEODIFF} caused by government censoring, or being removed by Google because of disruptive adware, malware, restricted content, or other reasons.
After these exclusions, we were left with 99,194 usable apps and their metadata. 
We then employed \textit{google play scraper}~\cite{googleplayscapper} to obtain app information including privacy policy links, app categories, and required device permissions.
Based on the available privacy policy links shown on the app's homepages, we further utilize the \textit{BeautifulSoup}~\cite{beautifulsoup} and \textit{Selenium}~\cite{Selenium} python packages to download those websites.
Eventually, we obtained 46,472 privacy policies, and we refer to this privacy policy dataset as \textbf{``Crawled Privacy Policy Collection.''}
We conducted the above data-gathering process in March 2022.


%
%
\begin{figure*}[t!]
  \centering
  \includegraphics[width=0.22\linewidth]{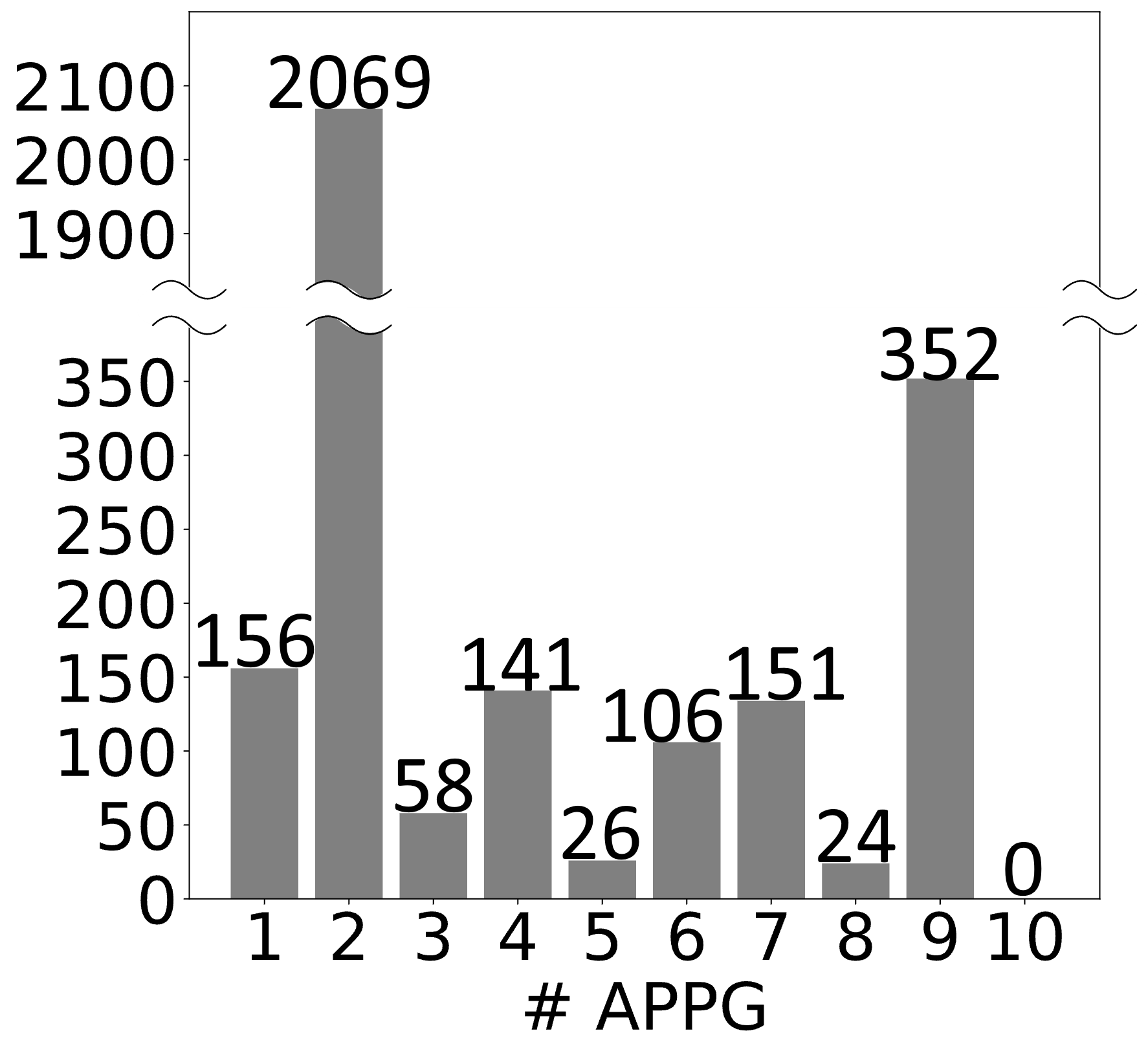}
  \includegraphics[width=0.22\linewidth]{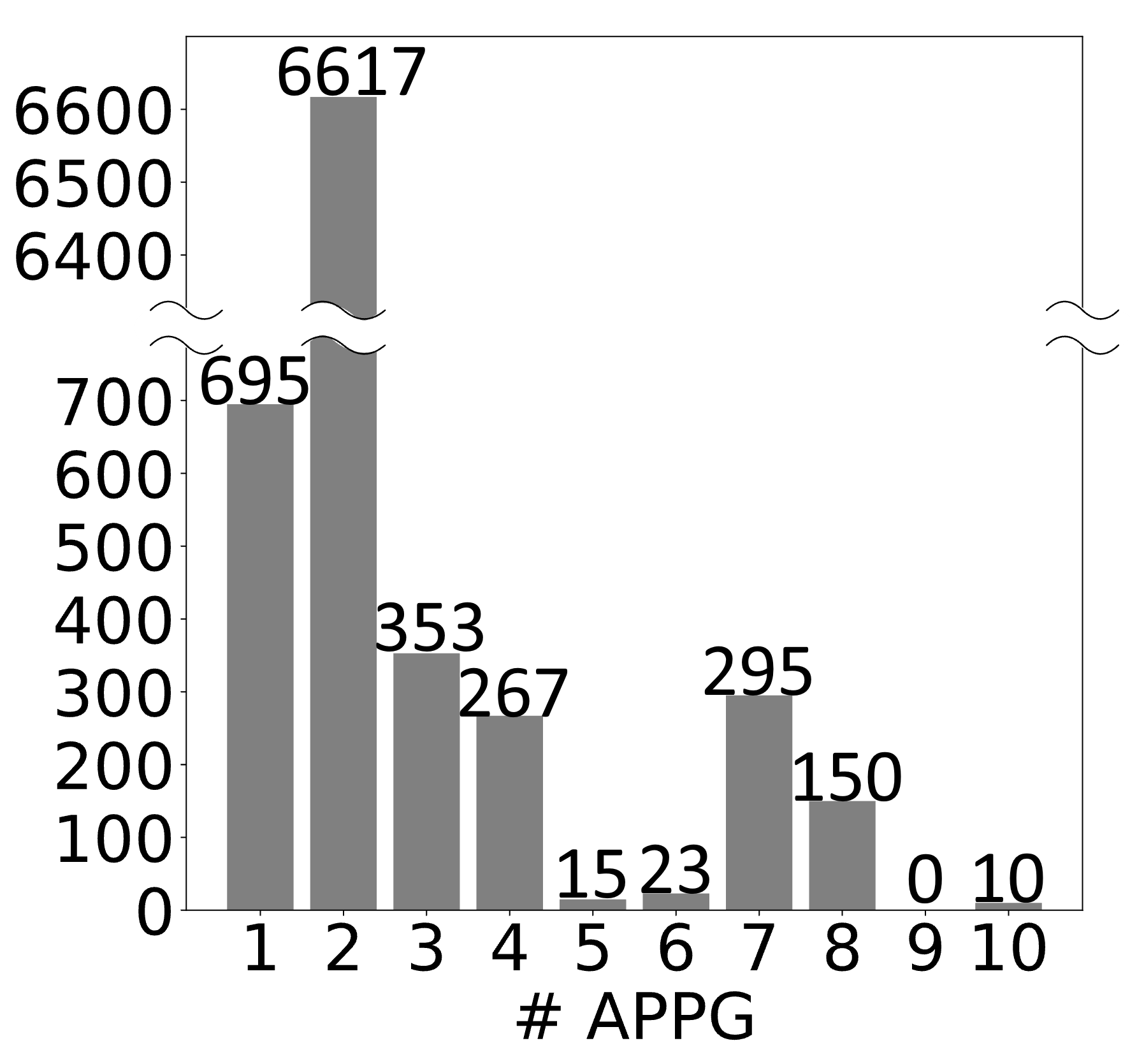}
  \includegraphics[width=0.22\linewidth]{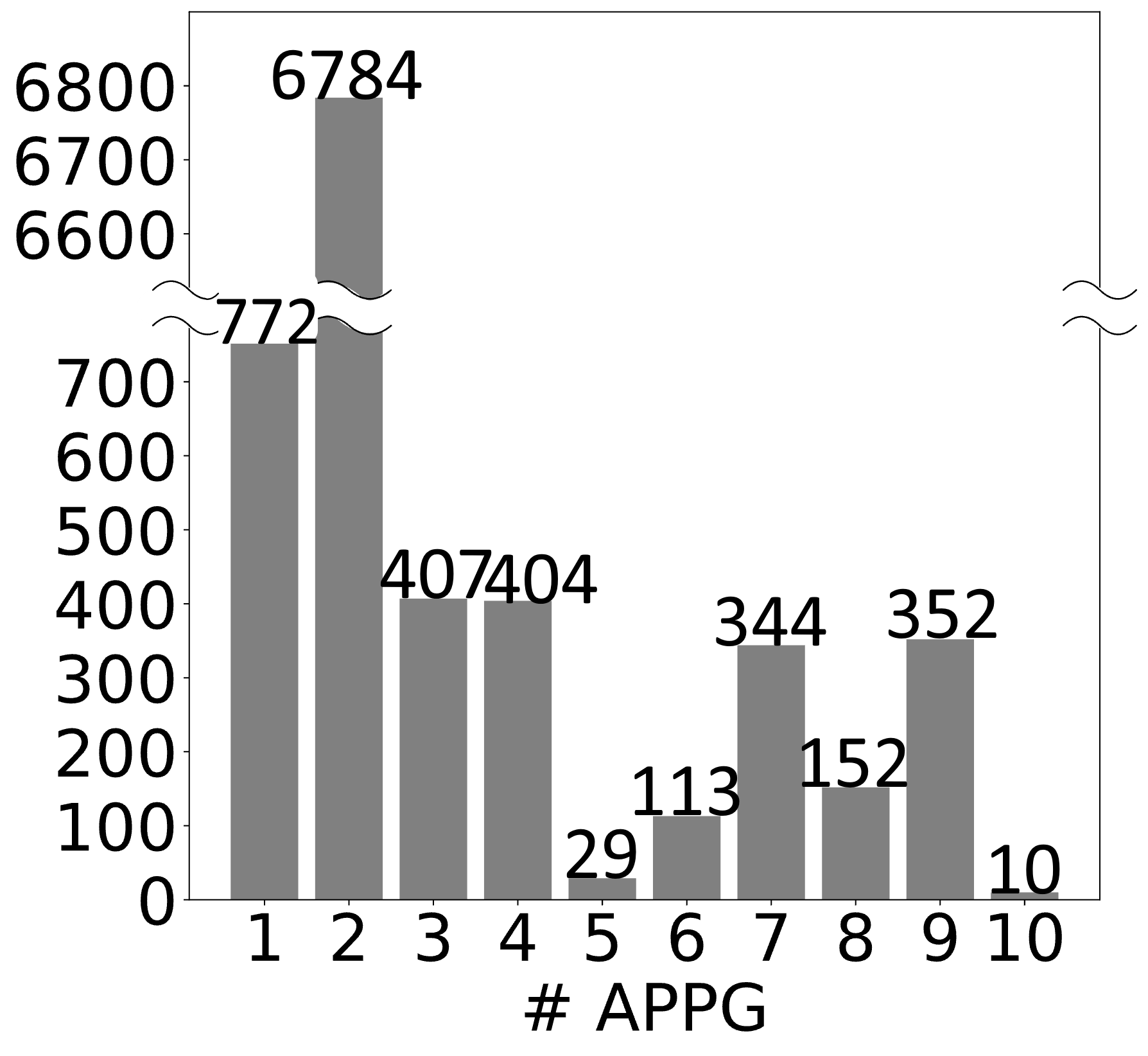}
  \caption{The number of privacy policies generated by APPGs, detected by FKS (left), DSC (mid), and both (right). }
  \label{fig_APPGcount}
\end{figure*}
%
%

\subsection{Observations on Status Quo}
\label{sec_overview_observation}
In this section, we report an analysis of app privacy policies and identify three common problems in existing apps: missing/unavailable privacy policies, low-quality privacy policies, and language non-localization problems.
Although similar problems were discussed in several previous studies~\cite{PETS_2019, USENIX_2022_GEODIFF, li2022understanding}, considering the rapid development of markets, we believe it is worthwhile to revisit those critical issues.

\subsubsection{Missing Links and Unavailable Privacy Policies}
\label{sec_observation_missing}

Providing a privacy policy is essentially required by platform principles and privacy regulations, however, we observed that 15.7\% (15,572/99,194) of apps do not provide a privacy policy link on their Google Play homepage.
A previous study~\cite{PETS_2019} reported that this statistic was 49.1\% in 2019. 
Seemingly, as more regulation has been introduced, more apps have provided a privacy policy.
However, we found that 37.5\% (37,150/99,194) of privacy policy links lead to unavailable websites with error messages such as ``403 Forbidden'' or ``404 Not Found.''
This problem could be caused by potential reasons such as deliberately providing a dummy link, an outage, or the removal of the website server.
Hence, in addition to simply requiring app developers to provide the privacy policy, the mobile app market also needs to frequently and regularly verify the validity of provided privacy policy links. Also, a permanent hosted privacy policy link provided by APPGs can help developers to mitigate availability issues to some extent.

\subsubsection{Low-Quality Privacy Policies}
\label{sec_observation_lowquality}


Several previous studies~\cite{liu2021have, PETS_2019} reported that it was not rare that some apps only provide a low-quality privacy policy, and we observe similar issues in our dataset.
Firstly, some privacy policies do not contain meaningful privacy-related context, or are just  dummy websites.
For example, ``Bway App'' is a free app that provides football results, statistics, trends, and match result predictions, with over 10,000 installs; however, their privacy policy website~\cite{BwayApp} does not contain any content, and the file size of the crawled privacy policy is close to 0 KB.
In addition, some are apparent ``homemade'' privacy policies that are too general to include essential data practices.
For instance, ``Easy Communication'' is designed to help people with autism, cerebral palsy, dyslexia, intellectual disability, and other special needs to communicate easily. Its privacy policy~\cite{EasyCommunication} only has 156 words, and ambiguously mentions required device sensor permissions without any elaborations.
Based on our observations and previous work, we empirically set the file size threshold as 2 KB~\cite{liu2021have}, and the document length threshold as 200 words~\cite{USENIX_2022_GEODIFF}. If a crawled privacy policy does not meet both criteria, it will be regarded as a low-quality privacy policy. In total, we identify 22.3\% (10,375/46,472) low-quality privacy policies.


\subsubsection{Language non-Localization Problem}
\label{sec_observation_language}
Language localization is necessary to promote apps in a global market. 
Although app UI and content can be translated, app developers may neglect the privacy policy. 
We employed the Python package \textit{langdetect}~\cite{langdetect} to detect the language of the crawled privacy policy documents and found 20.5\% (9,523/46,472) of apps do not provide an English privacy policy, although these apps were released in markets where the primary language is English. 
Interestingly, the top five non-English languages are Spanish (15.6\%), Portuguese (9.7\%), German (8.8\%), Korean (8.7\%), and French (7.6\%).
This trend might be attributable to the implementation of the European GDPR and Brazilian LGPD. 
We observed this problem even with top apps from some large companies. 
For the privacy policies of the top 1,000 most installed apps (all $>$ 500k installs) in our dataset, there are still around 14.1\% that do not provide an English privacy policy.
APPGs may be a solution to language non-localization problems. 
Only APPG \#1 \textit{Iubenda} supports multilingual generation, but some APPGs (\#4, \#7, and \#9) have included placeholders for future options for language selection.

\begin{table}[t!]
\caption{Summary of market use of different APPGs.}
\label{tab_market}
\centering
\resizebox{0.8\linewidth}{!}{%
\begin{tabular}{lr} 
\toprule
\textbf{Method}  & \textbf{Market Occupancy}\\
\midrule
Fingerprint Keyword Searching & 6.6\% (3,066) \\
Document Similarity Comparison & 18.1\% (8,425) \\
\midrule
Intersection &  4.4\% (2,042)\\
\textbf{Union (Total)} &  \textbf{20.1\% (9,332)}\\
\bottomrule
\end{tabular}
}%
\end{table}

\subsection{Synthetic Apps and Generated Privacy Policy Collection}
\label{sec_analysis_synthetic}

To identify whether privacy policies could be generated by APPGs, we first need to build the ground truth with self-generated privacy policies for each APPG. 
As shown in Table~\ref{tab_overview}, seven APPGs require users to register an account to use the service, and six of them also require users to pay subscription fees to unlock all features. 
We registered as required and paid subscription fees.
Synthetic apps are commonly used in similar empirical studies such as \cite{oltrogge2018rise, EASE_2020}.
We designed and tailored 3 synthetic apps specifically based on APPGs’ features and characteristics as summarised in Table~\ref{tab_characterization} and Table~\ref{tab_diversification}. This was so that the boilerplates and pre-defined clauses of every APPG can be fully explored and reflected in the generated privacy policies.
The functions and features of these synthetic apps are as follows:

\noindent \textbf{Synthetic App 1}. A toy-like app that collects only general personal information and does not need to comply with GDPR, CCPA, or LGPD. 
    
\noindent \textbf{Synthetic App 2}. A social app that enables people to interact and communicate with others. Users need to create an account by providing their general personal information, and the app requires all device permissions. The app also accesses third-party services and only needs to comply with GDPR. 
    
\noindent \textbf{Synthetic App 3}. A hypothetical omnipotent app that requires users to provide all general and sensitive personal information to function. The app also needs all device permission and access to third-party services. This app needs to meet the requirements of GDPR, CCPA, and LGPD.

These three synthetic apps are various in terms of functionality and sophistication, and they decently cover all data use mentioned in Table~\ref{tab_diversification}. 
Therefore, they are capable of being the boilerplate apps for APPGs' ground truth in the majority of cases.
For each APPG, we created the same three custom privacy policies to test whether a privacy policy can be generated by one of analysed APPGs.
Given ten APPGs and three synthetic apps, we obtained 30 privacy policies as \textbf{``Generated Privacy Policy Collection''} in total. We utilized it and ``Crawled Privacy Policy Collection'' to conduct the following market penetration analysis.

\subsection{Market Penetration Analysis}
\label{sec_analysis_market}

We report a market penetration analysis to highlight the extent to which APPGs are used and to provide insights about why some APPGs are more popular.
Specifically, we employ two methods to detect whether a privacy policy could be created by one of these APPGs: fingerprint keyword searching and document similarity comparison.

\textbf{Fingerprint Keyword Searching (FKS)}.
APPGs offer various publishing features, including the provision of a direct link as a permanent website host or an editable HTML document. 
We observe that to better advertise themselves, APPGs typically embed their company names into the URLs of the generated privacy policy websites or the editable HTML document. Consequently, we construct the fingerprint keyword set grounded in this feature.
To cultivate this set, we manually inspected these APPGs and the collection of generated privacy policies.
To mitigate the likelihood of false positives, we verified the authenticity of all keywords individually. Specifically, we incorporated a candidate fingerprint keyword into the set only if it appeared consistently across all three ground-truth privacy policies crafted for synthetic apps.
All keywords are deliberately sensitive to case and format.
The full keyword list is available at~\cite{our_repo}.

Then, the collected keyword set is employed to perform FKS in the crawled privacy policy collection.
For each privacy policy in our collection, if we find a match in either its website link or its document context with one of the fingerprint keywords, the policy is considered to be generated by the corresponding APPG. 
For example, during the examination of policy website URLs, we use the keyword ``iubenda.com'' to search URLs like "www.iubenda.com/123abc..." for APPG \#1 \textit{Iubenda}. 
Similarly, while scrutinizing policy HTML documents, identifiable phrases like "iubenda hosts this content" were utilized as markers


\begin{figure}[t!]
  \centering
  \includegraphics[width=1.0\linewidth]{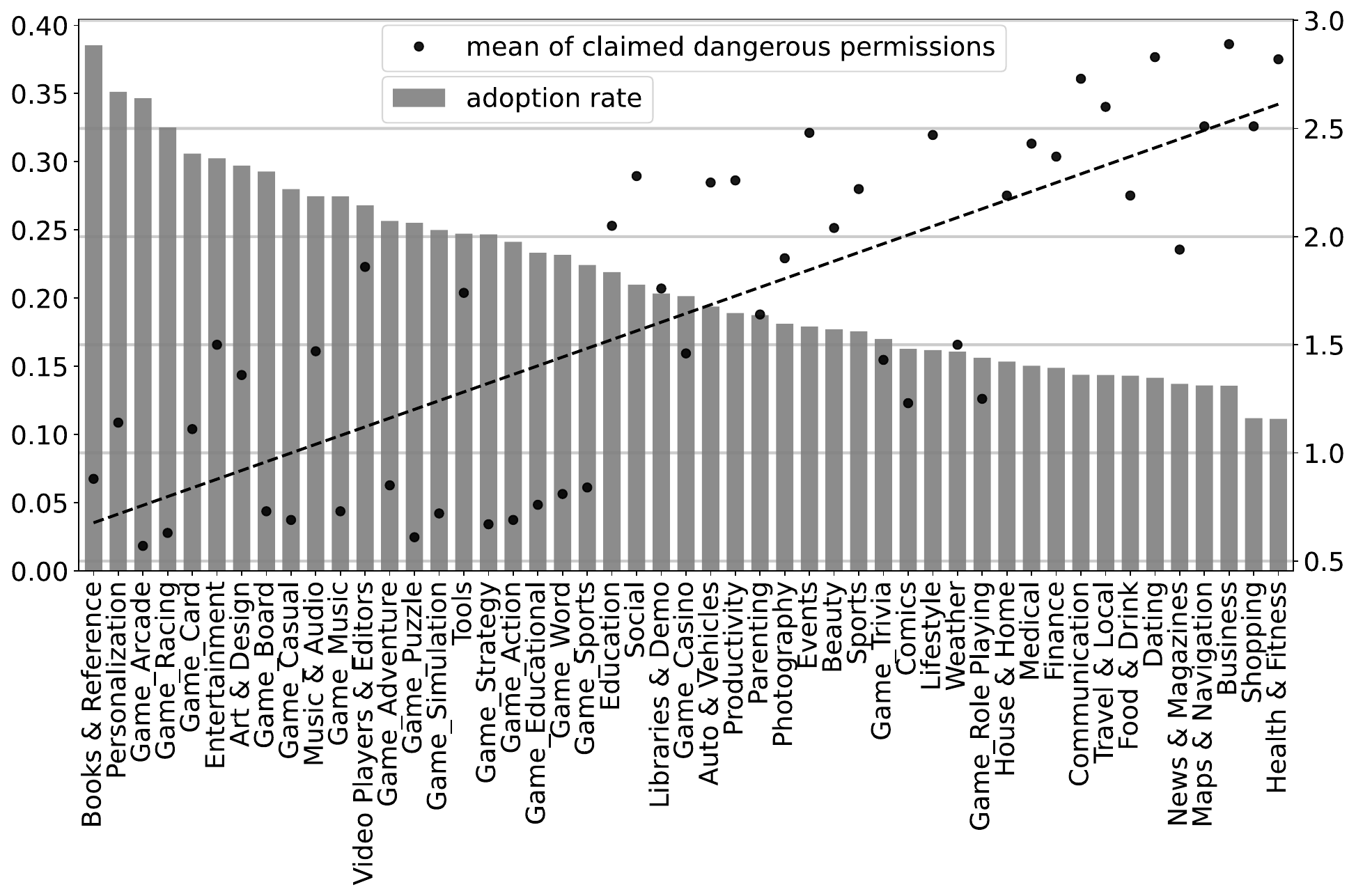}
  \caption{Use in market by APPGs (bars) and the average device permissions claimed (dots), by app categories. The left and right y-axis ticks represent the market occupancy percentages and the number of average permissions, respectively. The dashed line is a linear regression computed from the means of claimed dangerous permissions for each category.}
  \label{fig_categoryratio}
\end{figure}


\textbf{Document Similarity Comparison (DSC)}.
As we mention in Section~\ref{sec_characterization}, some APPGs allow users to customize the content to some extent, and APPG users may also further adapt or polish the initial generated privacy policy in editable HTML format before publishing. Consequently, the fingerprint keywords, even sentence-level segments, are not necessarily included in a generated app privacy policy sometimes, and so the searching-based method may not work in those cases. To tackle this problem, we design a semantic-based method with better robustness, called document similarity comparison.

In natural language processing (NLP) tasks, researchers adapt the concept from mathematics by treating documents as vectors, to obtain the similarity between two documents. 
TF-IDF~\cite{TFIDF} is a common numerical statistic designed to reflect the relative importance of words in a document corpus, and thus it can be adapted to convert documents into vectors.
By doing so, the similarity between two documents can be represented as the similarity of two vectors, and we choose cosine similarity as the most common one.
If the average similarity between a crawled privacy policy and three ground-truth generated privacy policies is greater than a threshold $\beta$, we consider it a match, i.e., could be created by the current APPG.
If a privacy policy matches more than one APPG, we take the one with the highest similarity.
To define the value of $\beta$, we first calculate the document similarities between the three generated privacy policies of each APPG as an initial value. We then randomly sampled 100 privacy policies that are generated by one of 10 APPGs based on fingerprint keyword searching and evaluate the similarities compared to the three generated privacy policies of each APPG to fine-tune the $\beta$. Based on our data, we empirically set $\beta = 0.75$.

\textbf{Results.}
We employed the proposed methods to provide an estimation of the market penetration of APPGs.
Among 46,472 valid privacy policy documents from apps at the Google Play Store, as shown in Table~\ref{tab_market}, we found 6.6\% (3,066) privacy policies are highly likely generated by one of the APPGs by using fingerprint keyword searching and 18.1\% (8,425) by using document similarity comparison.
We also checked the intersection size, which is 4.4\% (2,042), indicating that these two methods are complementary to each other.
The union of the previous two methods is \textbf{20.1\%} (9,332), indicating an upper bound on possible market occupancy ratio.
Overall, the results show that APPGs do play a considerable role in the current mobile app market.

Figure~\ref{fig_APPGcount} presents the market occupation  for each APPG.
We can see that APPG \#2 \textit{App Privacy Policy Generator} dominates as the most popular APPG, with a 72.7\% adoption rate. It could be attributed to its free-of-charge and registration, low complexity, and best readability score.
\#1 \textit{Iubenda} and \#5 \textit{App Privacy Policy} also take decent market shares. \#10 \textit{Lawpath} is the least used APPG in our data, perhaps related to its price to use of \$199 for one privacy policy document, which is the highest among all APPGs.
We notice popular APPGs generally have a low level or partial support on all data practice dimensions, therefore, we conclude that users may prefer to select easy-to-use tools rather than spend extra time learning how to master a sophisticated but more functional APPG. 
However, the ease of use of APPGs may come at the cost of a potentially-higher risk to breach privacy regulations.

%
%
\begin{figure}[t!]
  \centering
  \includegraphics[width=1.0\linewidth]{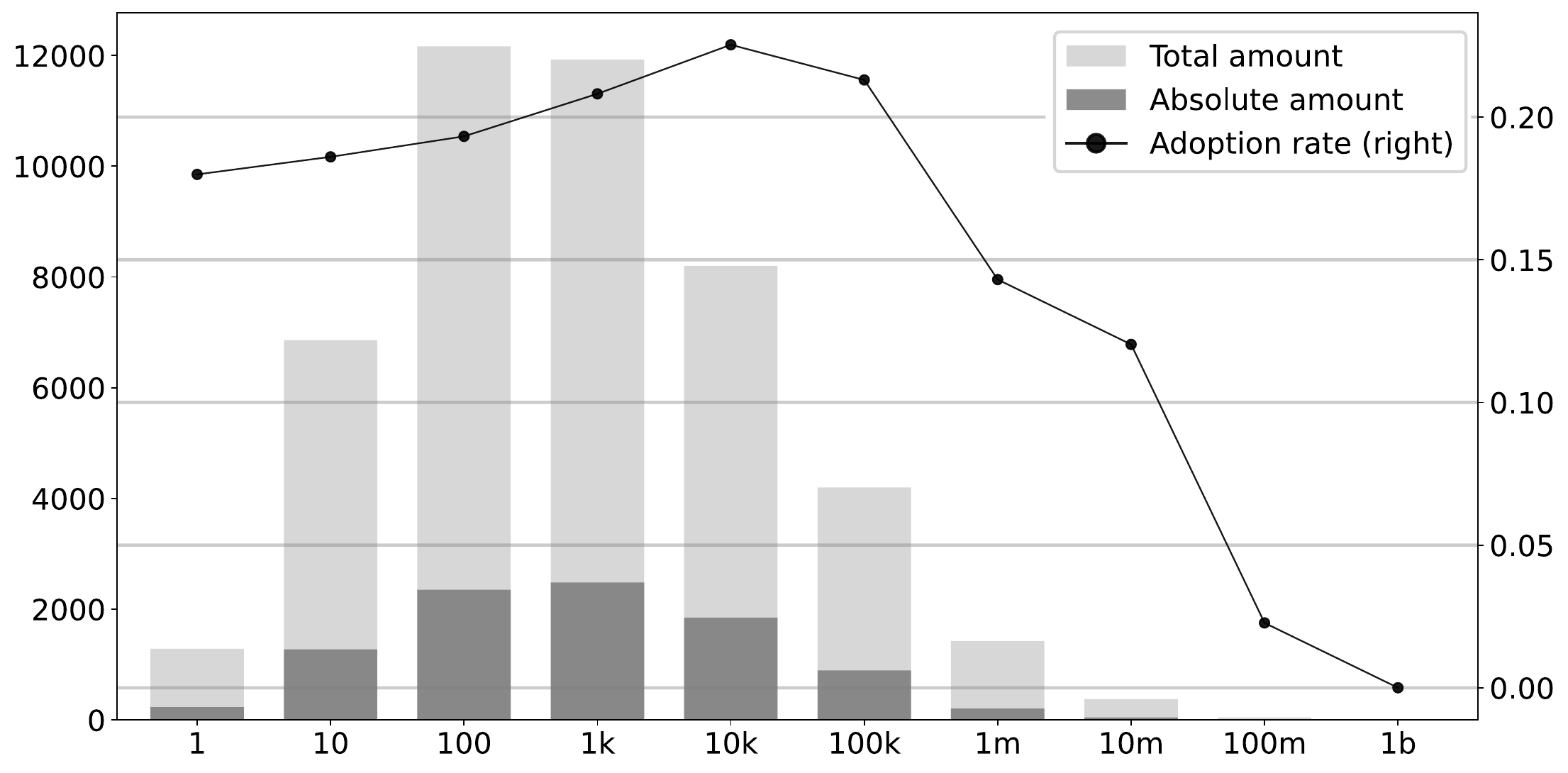}
  \caption{The distribution of app installs for all apps (light gray bars), the distribution of app installs for whose privacy policy could be generated by APPGs (gray bars), and the adoption rate for different install intervals (black dots). The installs mean the minimum install number an app has. ``1k'' denotes 1,000, ``1m'' denotes a million, and so on. The left y-axis ticks represent the number of apps, and the right y-axis ticks represent the adoption rate of APPGs.}
  \label{fig_installratio}
\end{figure}
%
%

\subsection{Trend of APPGs in Market}

This section describes a follow-up analysis aiming to reveal what kind of apps are more likely to employ APPGs in the market.
Figure~\ref{fig_categoryratio} reveals the relation between the adoption rate and the average device permissions, by app categories.
Specifically, developers tend to more commonly use APPGs to generate their privacy policies, for app categories with low average device permissions such as \textit{Books \& Reference} and game categories.
For apps categories that are prone to require more device permissions to function, for example, \textit{Health \& Fitness}, \textit{Shopping}, and \textit{Business}, their privacy policies tend not to be created by APPGs.

Figure~\ref{fig_installratio} shows the number of times apps have been installed and their APPGs' adoption rate. Apps tend to use APPGs when their installations increase, however, after 10k installations, the APPGs' adoption rate starts to rapidly drop. This trend confirms our observation that developers of less popular apps, i.e., those with fewer installations, tend to employ APPGs to quickly generate privacy policies to satisfy the requirements of regulations as well as the needs of end-users. However, developers of popular apps usually have more commercial resources to use manual alternatives and may have greater concerns about privacy practices because more users may increase privacy risks or attention from regulators.

\begin{tcolorbox}
     \textbf{Finding 2:} The market occupancy ratio of 10 examined APPGs is around 20.1\%, and \#2 \textit{App Privacy Policy Generator} is the most popular one, boasting a 72.7\% adoption rate. Moreover, users tend to select easy-to-use APPGs even though at the cost of a potentially-higher risk to breach privacy regulations.
\end{tcolorbox}

%% file: 5_Analysis.tex
\section{Assessment of Privacy Policies (\textbf{RQ3})}
\label{sec_analysis}
Generators are designed to generate compliant privacy policies. This section reports a detailed assessment of generated privacy policies.

%
\input{Tables/regulation}
\input{Tables/LGPD}
%

\subsection{Privacy Compliance Against Regulations}
\label{sec_analysis_regulation}

APPGs need to be designed in a way that allows developers to produce compliant privacy policies. Furthermore, they must be updated to reflect evolving laws. 
In Table~\ref{tab_characterization}, six APPGs claim that they are able to generate policies in compliance with CCPA and GDPR, and only \#1 \textit{Iubenda} mentions compliance with LGPD.
However, their actual compliance with generated policies may be different from what they claim.
Zimmeck et al.~\cite{zimmeck2021privacyflash} conducted two studies in May 2020 and January 2021, revealing that some APPGs had significant compliance issues, such as failing to create CCPA-compliant policies. 
We perform a follow-up inspection in May 2022 based on generated privacy policies by Synthetic App 3, since they are supposed to comply with all compliance requirements of GDPR, CCPA, and LGPD. 
Table~\ref{tab_regulation} shows tallies of generators’ compliance with legal requirements in May 2020, January 2021, and May 2022 for GDPR, CCPA, and LGPD.\footnote{The compliance analysis of GDPR and CCPA of May 2020 and Jan 2021 are directly obtained from~\cite{zimmeck2021privacyflash}. The specific requirements and compliance checking for GDPR and CCPA are tallied in Appendix Table~\ref{tab_gdpr} and Table~\ref{tab_ccpa}.} 
The specific requirements and compliance checking for LGPD are tallied in Table~\ref{tab_lgpd}. 

There is a notable discrepancy between the APPGs' claimed ability to generate policies compliant with specific regulations and their actual compliance levels.
In addition, across the table, compliance with GDPR and CCPA requirements improved over time from May 2020 to May 2022 for several APPGs (\#3, \#4, \#7, \#9), demonstrating an adaptation to regulatory standards.
An encouraging observation is that APPGs (\#1, \#3, \#4, \#7, \#8, \#9) maintained full GDPR and a majority of CCPA compliance since January 2021.
This might be due to the fact that GDPR has been in effect since 2018, providing APPGs with more time to adapt their policies accordingly.
Given that LGPD came into effect in September 2020, it is not unexpected that compliance levels would be relatively low in the initial stages.
The findings underscore the necessity for APPGs to improve their ability to produce compliant privacy policies consistently across all major regulations.
The results also call for ongoing monitoring and evaluation of APPGs to ensure they keep pace with evolving laws and maintain transparency with their users regarding their actual capabilities.

APPGs have varying levels of compliance when it comes to data collection practices. For instance, while GDPR emphasises "Categories of personal data concerned", CCPA requires to provide a detailed list of categories of personal information that should be disclosed if collected. APPGs \#1, \#3, \#4, and \#7 demonstrate compliance with GDPR's requirements, but when it comes to CCPA's detailed disclosure requirements, only APPGs \#3, \#7, and \#9 comply.
As for third-party data-sharing and data-sale practices, CCPA's emphasis is on "List of categories of personal information collected and sold" highlights this; whilst GDPR identifies ``Recipients or categories of recipients of the personal data''. APPGs \#1, \#3, \#4, and \#7 have clear disclosures in line with third-party disclosure, but others like \#2 and \#5 lag behind.
Overall, APPGs \#3 and \#7 consistently fare well across all three regulations, indicating a comprehensive privacy policy generation. However, APPGs \#2, as the most popular, display apparent gaps compared to others.

\textbf{Data Right Coverage Analysis.} Regulations often necessitate data rights to grant sufficient choice and control to the end users over the personal data collected by businesses. 
Thus, it is important to assess whether the generated policies clearly disclose data rights.
For this, following [GDPR Art.13] and [CCPA \textsection{}1798.130(a)(5)(A)]~\footnote{The specifications are stated in \textsection{}1798.100, \textsection{}1798.105, \textsection{}1798.106, \textsection{}1798.110, \textsection{}1798.115, and \textsection{}1798.125.}, we extracted and summarised seven fundamental data rights.
We manual examined the existence of disclosure for each data right of 10 APPGs. 
For comparison, we also scrutinized 12 leading mobile apps, which are mentioned in Section~\ref{sec_characterization}-Readability.
These 12 leading apps are in various categories and published by different companies, with over one billion installations.
Table~\ref{tab_dataright} shows that almost all leading apps disclose fundamental data rights in their privacy policies.
In addition, results validate the respective assertions made by APPGs, regarding their self-claimed compliance with GDPR and CCPA.
Notably, \#2 as the most popular APPG, does not disclose any of the data rights in the generated policy document.

%
\input{Tables/data_right}
%

\textbf{Privacy Practice Disclosure Analysis.}  
In alignment with the principles of data protection and transparency, it is also imperative that privacy policies encompass disclosure of privacy practices.
Based on \cite{googledeveloper} and \cite{SeeNoEvil}, this study delineates four highly concerning privacy practices. These are: 1) \textit{Data Encryption}: Users' data are encrypted and transferred over a secure connection. 2) \textit{Government Requests}: It refers to the potential for government entities to request access to users' data for various reasons, including national security or criminal investigations. 3) \textit{Data Breach Notification}: In the event of a data breach where users' information is compromised, it is essential to have a mechanism in place to notify affected users promptly. 4) \textit{Changes Notification}: The privacy policy should mention the procedures for notifying users about significant changes to the policy, including how and when users will be informed.
Then, we manually scrutinized the existence of disclosure across 10 APPGs and the same 12 leading apps. Table~\ref{tab_datapractice} shows that more than half APPGs disclose \textit{Data Encryption} and \textit{Changes Notification}. 
Five APPGs include \textit{Government Requests} and only \#8 includes \textit{Data Breach Notification} in the generated policy document (\textit{``In the event of a data breach, we will make reasonable efforts to notify affected individuals...''}).
In addition, \#5 and \#6 do not include any privacy practice, and \#8 includes all items.
Furthermore, \#2 as the most popular APPG, only include \textit{Data Encryption}.

%
\input{Tables/data_practice}
%

\begin{tcolorbox}
     \textbf{Finding 3.1:} Noncompliance with privacy laws and under-claiming issues indicate potential need for a more stringent evaluation of APPGs' capabilities.
\end{tcolorbox}

\subsection{Permissions Coverage Analysis}
\label{sec_analysis_permissions}

Device (sensor) permissions enable features of mobile apps and are normally critical in terms of personal privacy risks.
Intuitively and legally, an app's privacy policy should accurately disclose its permission usages; however, it is common that developers do not list all the permissions of their app in the privacy policy~\cite{andow2019policylint, PETS_2019, cui2023poligraph, pan2023seeprivacy, pan2023toward}.
Developers of mobile apps in the Google Play app store are required to clearly list the permissions they intend to obtain on their homepages\footnote{This feature was deprecated as Google launched the new Data Safety after we completed this work.}, and we denote them as $P_\text{claimed}$.
By taking the self-reported device permission usage as the ground-truth, we then assess whether the APPG's provided UIs, questionnaires, or documents can sufficiently allow the APPGs users to enter permission invocation information by their app. 
Notably, the self-reported device permission usages by developers may not actually reflect the actual privacy behaviour of apps.

To scrutinise how developers translate used device permissions into policies by APPGs, for each permission ($per$), we use it and Synthetic App 1 to generate a privacy policy ($PP_{App1 + per}$).
By comparing the differences between $PP_{App1 + per}$ and $PP_{App1}$, we can locate and obtain the phrases/sentences each APPG uses to declare each permission in the privacy policy.
For each app, by searching and counting the corresponding phrases/sentences in its privacy policy, we obtain the number of permissions displayed and denoted as $P_\text{displayed}$.
For each app that uses the APPG, its $P_\text{recognised}$ is equal to the recognised dangerous permissions enumerated in Table~\ref{tab_diversification}.
Then, we introduce two metrics for the extent to which APPGs cover claimed permissions in the generated privacy policy, as follows:
\begin{equation*}
\begin{aligned}
     \text{Recognised Coverage} \,(RC) = \frac{1}{n} \sum_1^n \frac{P_{\text{displayed}}}{P_\text{recognised}}
\end{aligned}
\end{equation*}
A higher $RC$ indicates the app developers use the APPG more properly, despite APPGs' intrinsic design flaws. And
\begin{equation*}
\begin{aligned}
     \text{Expected Coverage} \,(EC) = \frac{1}{n} \sum_1^n \frac{P_\text{displayed}}{P_\text{claimed}}
\end{aligned}
\end{equation*}
A higher $EC$ reflects the APPG has less intrinsic design flaws about disclosing permissions in generated privacy policies.

\input{Tables/permission_results}

We use the intersection of privacy policies identified by both FKS and DSC (2,042 privacy policies) because it is more certain that they are generated by one of the APPGs.
Based on our observations in Section~\ref{sec_diversification}, four APPGs (\#2, \#5, \#6, \#10) do not support permission declarations, in other words, $P_\text{recognised} = 0$ for every app identified to use those APPGs.
In addition, we find that $P_\text{displayed} = 0$ for four of those apps, which means that none of the developers realised that the APPG they used did not cover the permissions they claimed.

The results are shown in Table~\ref{tab_permission}.
First, for \#1 and \#3, they both can recognise all nine device permissions involved in this study, so their $RC$s are equal to $EC$s.
There is a significant portion of missing permission disclosure for all APPGs, since all $EC$s are much smaller than 100\%.
The missing disclosure issue is not dependent on users' input, but on the design of APPGs that did not sufficiently accommodate the requirements.
Second, compared to \#1's and \#3's consistency on $RC$s and $EC$s, for \#4, \#7, \#8, and \#9, their $EC$s are greatly smaller than $RC$s.
The gap shows the importance of APPG selection, because inappropriate APPGs will hinder developers from including self-reported (claimed) permissions in their privacy policies.
Third, \#1 has the lowest $RC$, but it has high level user instruction; whereas \#4, \#7, and \#9 has the relatively high $RC$s, but low level of user instruction.
This counter-intuitive result indicates that current user instructions are not helpful enough to guide developers to correctly include permissions in the generated policies.
In addition, \#1 is the only UI-mode APPG, with the lowest $RC$ (15.6\%), and other comparable APPGs, whose $RC$ is over 50\% are all questionnaire-mode.
This observation indicates that UI mode APPGs may be more challenging to be properly used by users.

\begin{tcolorbox}

     \textbf{Finding 3.2:} A significant portion of device permissions remain under-claimed, perhaps caused by APPGs' design issues, and questionnaire mode can better guide users to include claimed device permissions compared to UI mode.

\end{tcolorbox}

\subsection{Contradiction Analysis}
\label{sec_contradiction}

Previous works have discussed contradictions within a policy~\cite{andow2019policylint, cui2023poligraph}. Specifically, if affirmative and negative sentences mention the same or conflicting entities and data types in a policy, then there can be a contradiction.
In this section, we compare the APPG-based and non-APPG-based privacy policies, to examine the situation of contradiction issues in both and the effect introduced by APPGs. 
For non-APPG-based privacy policies, we used the test set from PoliCheck~\cite{andow2020actions} which contains 200 policies randomly sampled from 13K mobile apps.
For APPG-based privacy policies, we use 30 policies from the ``\textbf{Generated Privacy Policy Collection}'' and 170 policies that are randomly sampled from the intersection list (indicated as APPG-generated by both FKS and DSC).
We employ the state-of-the-art privacy policy analyser, PoliGraph~\cite{cui2023poligraph}, to conduct the following analysis.

We observed \textit{26} and \textit{15} contradictions for APPG-based and non-APPG-based privacy policies, respectively, suggesting that APPGs might introduce more privacy policy contradiction issues.
A detailed evaluation revealed that a significant proportion of these contradictions within APPG-based policies originate from conflicting statements found in separate sections of the document.
For example,  one section of the policy explicitly states ``[A condition], we sell your personal information to third parties'', while an assertion in the CCPA section contradicts this by claiming ``[The App] has not disclosed or sold any personal information to third parties for a business or commercial purpose in the preceding 12 months.''.
Therefore, APPG users should be vigilant in critically evaluating coherence and consistency  between different sections within privacy policies.

\begin{tcolorbox}

     \textbf{Finding 3.3:} More privacy policy contradiction issues exist in APPG-based privacy policies.

\end{tcolorbox}

%% file: Tables/regulation.tex
\begin{table}[t!]
\caption{Tallies of the APPGs’ compliance against legal requirements in privacy regulations. The individual requirements of LGPD are shown in Table~\ref{tab_lgpd}. ``N.R.'' stands for ``no record''. The enforcement date of LGPD is September 2020.}
\label{tab_regulation}
\centering
\resizebox{0.9\linewidth}{!}{%
\begin{tabular}{c|c|c|c} 
\toprule

\textbf{\#} & \begin{tabular}[c]{c@{}c@{}}\textbf{GDPR}\\\text{May'20  Jan'21  May'22}\end{tabular} & \begin{tabular}[c]{c@{}c@{}}\textbf{CCPA}\\\text{May'20  Jan'21  May'22}\end{tabular} & \begin{tabular}[c]{c@{}c@{}}\textbf{LGPD}\\\text{May'22}\end{tabular}  \\ 
\midrule
1                                                 
& 8/8 \hspace{15pt} 8/8  \hspace{15pt}  8/8        
& 14/18 \quad 14/18 \quad  14/18                    
& 8/8                        
\\
\midrule
2
& N.R. \hspace{9pt} N.R.  \hspace{9pt} 3/8
& N.R. \hspace{12pt} N.R.  \hspace{12pt} 3/18          & 3/8     
\\
\midrule
3
& 8/8 \hspace{15pt} 8/8  \hspace{15pt}  8/8 
& \hspace{1pt} 3/18 \hspace{10pt} 15/18 \hspace{9pt} 15/18          & 6/8     
\\
\midrule
4                                                    & 8/8 \hspace{15pt} 8/8  \hspace{15pt}  8/8 
&  \hspace{0.5pt} 5/18 \hspace{9.4pt} 16/18  \hspace{9.4pt} 16/18         
& 6/8                                                 \\
\midrule
5 
& N.R. \hspace{9pt} N.R.  \hspace{9pt}   0/8 
& N.R. \hspace{12pt} N.R.  \hspace{12pt} 2/18         & 1/8                                                 \\
\midrule
6 
& N.R. \hspace{9pt} N.R.  \hspace{9pt}   0/8 
& N.R. \hspace{12pt} N.R.  \hspace{12pt} 2/18         & 1/8                                                 \\
\midrule
7                                                    & 8/8 \hspace{15pt} 8/8  \hspace{15pt}  8/8  
&  \hspace{0.5pt} 5/18 \hspace{9.4pt} 16/18  \hspace{9.4pt} 16/18         
& 6/8                                   \\
\midrule
8                                                    & N.R. \hspace{9pt} N.R.  \hspace{9pt} 8/8 
& \hspace{1pt} N.R. \hspace{12pt} N.R.  \hspace{10pt} 11/18 
& 5/8                    \\
\midrule
9                                                    & 8/8 \hspace{15pt} 8/8  \hspace{15pt}  8/8  
&  \hspace{0.5pt} 5/18 \hspace{9.4pt} 16/18  \hspace{9.4pt} 16/18 
& 6/8                                   \\
\midrule
10                                                   & N.R. \hspace{9pt} N.R.  \hspace{9pt}   2/8 
& N.R. \hspace{12pt} N.R.  \hspace{12pt} 2/18   
& 4/8                    
\\
\bottomrule
\end{tabular}
}%
\end{table}

%% file: Tables/LGPD.tex
\begin{table*}[tbp]
\centering
\caption{LGPD privacy policy requirements and APPGs’ compliance in May 2022. Numbers indicate APPGs as per Table \ref{tab_overview}.}
\label{tab_lgpd}
\resizebox{0.85\textwidth}{!}{%
\begin{tabular}{l |ccccc| ccccc} 
\toprule
\textbf{LGPD requirement} 
&1 &2 &3 &4 &5 
&6 &7 &8 &9 &10\\ 
\midrule
\makecell[l]{Identify and document the legal bases for processing personal data. [Art. 1, 2]}
& \cmark
& \xmark
& \cmark
& \cmark
& \xmark
& \xmark
& \cmark
& \cmark
& \cmark
& \cmark
\\
\midrule
Disclose, collect and maintain valid proof of consent. [Art. 8 \textsection{1-6}] 
& \cmark
& \xmark
& \xmark
& \xmark
& \xmark
& \xmark
& \xmark
& \cmark
& \xmark
& \xmark
\\
\midrule
\makecell[l]{Include required data collection disclosures in the privacy policy in a clear, \\adequate, and notable manner. [Art. 9]}
& \cmark
& \cmark
& \cmark
& \cmark
& \xmark
& \xmark
& \cmark
& \cmark
& \cmark
& \cmark
\\
\midrule
\makecell[l]{Disclose the essential data rights to the data subject in a clear, adequate and \\ostensible manner. [Art. 9, 19]}
& \cmark
& \xmark
& \cmark
& \cmark
& \xmark
& \xmark
& \cmark
& \cmark
& \cmark
& \xmark
\\
\midrule
\makecell[l]{Appoint a data protection officer (DPO) and publicly disclosed the contact\\ information [Art. 41 \textsection{1}]}
& \cmark
& \xmark
& \cmark
& \xmark
& \xmark
& \xmark
& \xmark
& \xmark
& \xmark
& \xmark
\\
\midrule
\makecell[l]{Implement and disclose a security protocol (or such mechanism) to protect \\ personal data. \hspace{1pt} [Art. 46]}
& \cmark
& \cmark
& \cmark
& \cmark
& \xmark
& \xmark
& \cmark
& \xmark
& \cmark
& \cmark
\\
\midrule
\makecell[l]{State and comply with cross-border data transfer requirements if the data \\controller conducts international transfer of personal data. [Art. 33]}  
& \cmark
& \xmark
& \cmark
& \cmark
& \xmark
& \xmark
& \cmark
& \xmark
& \cmark
& \cmark
\\
\midrule
\makecell[l]{Disclose the processing of children’s data in a specific and highlighted consent,\\ including types of data collected, the way it is used and data rights. [Art. 14]}
& \cmark
& \cmark
& \xmark
& \cmark
& \cmark
& \cmark
& \cmark
& \cmark
& \cmark
& \xmark
\\
\bottomrule
\end{tabular}
}%
\end{table*}

%% file: Tables/data_right.tex
\begin{table}[h!]
  \caption{The disclosure existence of seven fundamental data rights. Numbers in the first row indicate APPGs as per Table \ref{tab_overview}, and ``Apps'' denotes the tallies of disclosure for 12 leading apps.}
  \label{tab_dataright}
  \centering
  \resizebox{1.0\linewidth}{!}{%
  \begin{tabular}{l|ccccc |ccccc ||c}
    \toprule
    \textbf{Data Right} & 1 & 2 & 3 & 4 & 5 & 6 & 7 & 8 & 9 & 10 & Apps \\
\midrule
Right to Know &\cmark &\xmark &\cmark &\cmark       &\xmark &\xmark &\cmark &\cmark &\cmark&\cmark & 12/12\\
\midrule
     Right to Access &\cmark &\xmark &\cmark &\cmark &\xmark &\xmark &\cmark &\cmark &\cmark&\cmark& 12/12\\
\midrule
    Right to Processing &\cmark &\xmark &\cmark &\cmark &\xmark &\xmark &\cmark &\cmark &\cmark&\xmark& 12/12\\
\midrule
    Right to Restrict of Processing &\cmark &\xmark &\cmark &\cmark &\xmark &\xmark &\cmark &\cmark &\cmark&\xmark& 12/12\\
\midrule
    Right to be Forgotten &\cmark&\xmark &\cmark &\cmark &\xmark &\xmark&\cmark &\cmark &\cmark &\cmark& 12/12\\
\midrule
    Right to Data Transfer &\cmark &\xmark &\cmark &\cmark &\xmark &\xmark &\cmark &\cmark &\cmark&\xmark& 12/12\\
\midrule 
    Right to Lodge a Complaint &\cmark &\xmark &\cmark &\cmark &\xmark &\xmark &\cmark &\cmark &\cmark&\cmark& 12/12\\
  \bottomrule
\end{tabular}
}%
\vspace{-5pt}
\end{table}

%% file: Tables/data_practice.tex
\begin{table}[h!]
  \caption{The disclosure existence of five highly concerning privacy practices. Numbers in the first row indicate APPGs as per Table \ref{tab_overview}, and ``Apps'' denotes the tallies of disclosure for 12 leading apps.}
  \label{tab_datapractice}
  \centering
  \resizebox{1.0\linewidth}{!}{%
  \begin{tabular}{l|ccccc |ccccc ||c}
    \toprule
    \textbf{Privacy Practice} & 1 & 2 & 3 & 4 & 5 & 6 & 7 & 8 & 9 & 10 & Apps \\
\midrule
Data Encryption 
&\cmark & \cmark &\cmark &\cmark &\xmark 
&\xmark &\cmark &\cmark &\cmark&\cmark 
& 8/12\\
\midrule
Government Requests 
&\cmark &\xmark &\xmark &\cmark &\xmark 
&\xmark &\cmark &\cmark &\cmark&\xmark
& 11/12 \\
\midrule
Data Breach Notification 
&\xmark  &\xmark &\xmark &\xmark &\xmark 
&\xmark &\xmark &\cmark &\xmark&\xmark
& 1/12\\
\midrule
Changes Notification
&\cmark&\xmark &\cmark &\cmark &\xmark 
&\xmark&\cmark &\cmark &\cmark &\xmark
& 12/12\\
  \bottomrule
\end{tabular}
}%
\end{table}

%% file: Tables/permission_results.tex
\begin{table}[t]
\centering
\caption{Permissions coverage for APPGs.}
\label{tab_permission}
\resizebox{0.95\linewidth}{!}{%
\begin{tabular}{cl|cc|cc} 
\toprule

\textbf{\#}     & \textbf{APPG Name}                              & Mode  & User instruction  &  RC & EC \\ 
\midrule
1&Iubenda& UI &\CIRCLE & 15.6\% & 15.6\%  \\
3&Termly & Questionnaire &\CIRCLE & 60.3\% & 60.3\% \\
\midrule
\midrule
4&Privacy Policies&Questionnaire&\Circle                                                                                                                       & 85.6\%   & 40.1\%                                                                            \\

7&Terms Feed& Questionnaire&\Circle                                                                                                                            & 52.7\%      & 25.7\%                                                                     \\

8&Website Policies& Questionnaire&\LEFTcircle                                                                                                                      & 60.0\%      & 11.8\%                                                                     \\

9&Free Privacy Policy& Questionnaire&\Circle                                                                                                                   & 64.4\%       & 25.6\%                                                                        \\
\bottomrule
\end{tabular}
}%
\end{table}

%% file: 6_Discussion.tex
\section{Discussion and Implication}

Challenges and opportunities coexist in the current APPG development. Based on our observations and study results, we summarize some findings for various roles or stakeholders in the APPG ecosystem.

\noindent \textbf{App developers/APPG users}.
While app developers may benefit from using APPGs to create privacy policies more efficiently, they should be aware of APPGs' latent limitations.
As illustrated in Section~\ref{sec_APPG}, APPGs have different qualities.
Some do not directly target specific regulations, such as GDPR and CCPA, and some do not recognise data practices, i.e., the declaration of personal information, device permissions, and third-party services used in their mobile apps.
Since app developers must make sure the privacy policies they provide are comprehensive, readable, and compliant, \underline{\textbf{it is very likely a trap}} if they do not select and use appropriate APPG with careful consideration.

\noindent \textbf{APPG providers}. Our analysis suggests APPG providers should work on improving recognised data use, since the majority of APPGs on the market only provide a very limited scope of personal information and device permissions.
The following two requirements are commonly neglected by the APPGs even though they claim to comply with CCPA:
\textbf{a)} Special requirements for businesses buying, receiving, selling, or sharing personal information of
10,000,000 or more consumers in a calendar year [Regs \textsection{} 999.308(c)(8), 999.317(g)(1)], and
\textbf{b)} For online notices, follow generally recognized industry standards [Regs \textsection{}999.308(a)(2)(d)]. 
Based on our market penetration analysis, we found that developers would rather use a simpler APPG (e.g., \#2) than a correct one.
It is worthwhile to consider this trade-off in the design and acquisition of APPGs.
Our results indicate that current user instructions are not helpful enough; thus, APPG providers should work to improve user instructions and provide intuitive UIs to guide users to correctly use their tools.
We have disclosed our observations and findings to APPG providers~\cite{our_repo}.

\noindent \textbf{Privacy regulators.} We discovered that the privacy policy of 20.1\% apps could be generated by existing APPGs, which means APPGs can significantly contribute to reducing policy absence and low quality privacy policies in app markets.
However, our results also highlight that the APPGs' design flaws can limit the app developers to include all essential privacy practices, data rights, and device permissions in the policies, which in turn creates a substantial risk of breaching privacy regulations.
Regulators should recognize the importance of this issue and be engage with this emerging market trend.
One possible approach would be to proactively provide guidance for APPGs developers, to improve compliance with privacy regulation, and to protect apps developers from potential pitfalls.
Finally, we hope this study could raise broad attention to inherent design shortcomings and resulting privacy concerns of online low-code/no-code tools such as APPGs, as they are increasingly employed in practice.

%% file: 7_RelatedWork.tex
\section{Related Work}
\textbf{Privacy policy analysis.}
Privacy policies have been extensively analysed and discussed by the research community~\cite{ACL_2016, PETS_2019, amos2021privacy, andow2019policylint, bui2021consistency}. 
Wilson et al.~\cite{ACL_2016} created a corpus of 115 privacy policies and 23K fine-grained data practise annotations, and revealed users' preferences on privacy policy structure and complexity.
Amos et al.~\cite{amos2021privacy} reported that privacy compliance and readability were worse in the last 20 years, according to 130k website privacy policies.
Andow et al.~\cite{andow2019policylint} presented \textit{PolicyLint}, which is a privacy policy analysis tool that can identify privacy contradictions by simultaneously considering negation and varying semantic levels of data objects and entities.
Bui et al.~\cite{bui2021consistency} proposed an automated system, dubbed \textit{PurPliance}, that detects inconsistencies between the data-usage purposes stated in a privacy policy and the actual behaviours of an Android app.

\textbf{Code-centric privacy policy auto-generation tools.}
Yu et al.~\cite{yu2015autoppg, yu2018ppchecker} developed a system named \textit{AutoPPG}, to automatically construct readable descriptions from the source code of mobile apps, to help create privacy policies on Android.
Rocky Slavin~\cite{polidroid} developed \textit{PoliDriod}, an Android Studio plugin that can be used to detect possible misalignments between Android API methods and privacy policies.
Zimmeck et al.~\cite{zimmeck2021privacyflash} proposed a privacy policy generator, named \textit{PrivacyFlash}, which leverages mappings between code signatures and privacy practises expressed in policies for iOS apps.
While code-centric generators might better align with apps' actual privacy behaviours compared to online APPGs, they come with significant challenges that hinder their adoption by app developers. First, they cannot ensure compliance with high-level privacy regulations, particularly non-functional requirements. Second, they present a higher entry barrier for developers due to their inherent complexity, whereas online APPGs can be easily accessed, offering a range of user instructions and publishing support as demonstrated in this study.
Others~\cite{bateni2021automated, windl2022automating} also discussed the automated generation of privacy policies using machine learning methods.


%% file: 8_Conclusion.tex
\section{Conclusion}
\label{sec_conclusion}

Online Automated Privacy Policy Generators (APPGs) are broadly used by developers of mobile apps to create privacy policies to respond to regulatory requirements. 
This paper reports the first large-scale empirical study to comprehensively scrutinize APPGs' various features, characteristics, and extent of recognition of data use.
Our market penetration analysis indicates that privacy policies of 20.1\% apps could be generated by existing APPGs on the Google Play app store, and that \#2 is the most popular APPG, with
a 72.7\% adoption rate.
Our findings underline a substantial level of noncompliance with privacy laws and a frequent under-claiming of data rights and highly concerning privacy practices, especially with the most popular APPG \#2.
Permissions coverage analysis reveals that existing APPGs have significant issues with including all essential device permissions in the generated privacy policy, and UI mode APPGs could make it worse.
Also, more contradiction issues exist in APPG-based privacy policies.
In summary, for app developers, selecting and employing APPGs without careful consideration is very likely a trap, creating a substantial risk of breaching privacy regulations.

%% file: 10_Acknowledgement.tex
\section* {Acknowledgements}
We thank Zhen Tao, Bowen Xu, and Chen Gong for useful feedback and help on earlier versions of this paper.
We thank the anonymous reviews and the Shepherd for providing valuable comments and suggestions. 

%% file: 12_Appendix.tex
\section*{Appendix}

\subsection*{Readability Analysis}

The readability score of privacy policies are calculated by the Flesch Reading-ease Test~\cite{kincaid1975derivation}:
\begin{equation*}
\label{readability}
\begin{aligned}
    206.835-1.015\left(\frac{\text {total words}}{\text {total sentences}}\right)-84.6\left(\frac{\text {total syllables}}{\text {total words}}\right)
\end{aligned}
\end{equation*}
Specific readability scores are listed in the Table~\ref{tab_readability}.

%
%
\input{Tables/readability}
%
%

\subsection*{GDPR and CCPA requirements compliance}

The specific requirements and compliance
checking for GDPR and CCPA are tallied in Appendix Table~\ref{tab_gdpr} and Table~\ref{tab_ccpa}. (In next page)

%
%
\input{Tables/GDPR}
%
%

%
%
\input{Tables/CCPA}
%
%

%% file: Tables/readability.tex
\begin{table}[t!]
  \caption{Comparison of APPGs' readability scores.}
  \label{tab_readability}
  \centering
  \resizebox{0.42\textwidth}{!}{%
  \begin{tabular}{r|l|c}
    \toprule
    \textbf{\#} & \textbf{APPG Name} & \textbf{Average Readability Score} \\
    \midrule
    \rowcolor{lightgray!85}
    1 & Iubenda & 37.3 \\
    \rowcolor{lightgray!15}
    2 & App Privacy Policy Generator &\textbf{49.3}\\
    \rowcolor{lightgray!85}
    3 & Termly & 46.0\\
    \rowcolor{lightgray!15}
    4 &Privacy Policies &37.7\\
    \rowcolor{lightgray!85}
    5 &App Privacy Policy & 43.9\\  
    \rowcolor{lightgray!15}
    6 &Privacy Policy Online  & 44.0\\
    \rowcolor{lightgray!85}
    7 &Terms Feed  & 37.7\\
    \rowcolor{lightgray!15}
    8 &Website Policies & 36.7\\
    \rowcolor{lightgray!85}
    9 &Free Privacy Policy & 37.7 \\
    \rowcolor{lightgray!15}
    10 &Lawpath & 41.0\\
    \midrule
    \rowcolor{lightgray!85}
    / & 12 leading apps & 46.0\\
  \bottomrule
\end{tabular}%
}
\end{table}

%% file: Tables/GDPR.tex
\begin{table*}[t!]
\centering
\caption{GDPR privacy policy requirements and APPGs’ compliance in May 2022. Numbers indicate APPGs as per Table \ref{tab_overview}.}
\label{tab_gdpr}
\resizebox{0.9\textwidth}{!}{%
\begin{tabular}{l |ccccc| ccccc} 
\toprule
\textbf{GDPR requirement} 
&1 &2 &3 &4 &5 
&6 &7 &8 &9 &10\\ 
\midrule
\makecell[l]{Identity and contact details of the data controller and their representative, \\if any [Art. 13(1)(a), 14(1)(a)]}
& \cmark
& \xmark
& \cmark
& \cmark
& \xmark
& \xmark
& \cmark
& \cmark
& \cmark
& \xmark
\\
\midrule
\makecell[l]{Purposes of the processing for which the personal data are intended and legal \\ basis for the processing [Art. 13(1)(c), 14(1)(c)]}
& \cmark
& \xmark
& \cmark
& \cmark
& \xmark
& \xmark
& \cmark
& \cmark
& \cmark
& \cmark
\\
\midrule
\makecell[l]{Categories of personal data concerned [Art. 14(1)(d)]}
& \cmark
& \cmark
& \cmark
& \cmark
& \xmark
& \xmark
& \cmark
& \cmark
& \cmark
& \xmark
\\
\midrule
\makecell[l]{Recipients or categories of recipients of the personal data [Art. 13(1)(e), 14(1)(e)]}
& \cmark
& \xmark
& \cmark
& \cmark
& \xmark
& \xmark
& \cmark
& \cmark
& \cmark
& \cmark
\\
\midrule
\makecell[l]{Period for which the personal data will be stored, or if that is not possible, the \\ criteria used to determine that period [Art. 13(2)(a), 14(2)(a)]}
& \cmark
& \xmark
& \cmark
& \cmark
& \xmark
& \xmark
& \cmark
& \cmark
& \cmark
& \xmark
\\
\midrule
\makecell[l]{Existence of the rights to request data access, rectification, erasure, and data \\ portability as well as the rights to restrict and object to processing [Art. 13(2)(b), 14(2)(c)] }
& \cmark
& \cmark
& \cmark
& \cmark
& \xmark
& \xmark
& \cmark
& \cmark
& \cmark
& \xmark
\\
\midrule
\makecell[l]{Right to withdraw consent for processing at any time [Art. 13(2)(c), 14(2)(d)]}  
& \cmark
& \xmark
& \cmark
& \cmark
& \xmark
& \xmark
& \cmark
& \cmark
& \cmark
& \xmark
\\
\midrule
\makecell[l]{Right to lodge complaint with a supervisory authority [Art. 13(2)(d), 14(2)(e)]}
& \cmark
& \cmark
& \cmark
& \cmark
& \xmark
& \xmark
& \cmark
& \cmark
& \cmark
& \xmark
\\
\bottomrule
\end{tabular}
}%
\end{table*}

%% file: Tables/CCPA.tex
\begin{table*}[t!]
\centering
\caption{CCPA privacy policy requirements and APPGs’ compliance in May 2022. Numbers indicate APPGs as per Table \ref{tab_overview}.}
\label{tab_ccpa}
\resizebox{1.0\textwidth}{!}{%
\begin{tabular}{l |ccccc| ccccc} 
\toprule
\textbf{CCPA requirement} 
&1 &2 &3 &4 &5 
&6 &7 &8 &9 &10\\ 
\midrule
\makecell[l]{Disclosure of right to request how personal information is collected, used, sold, disclosed for a business \\purpose, and shared [CCPA \textsection{}1798.130(a)(5)(A), 1798.110(a), 1798.115(a), Regs \textsection{} 999.308(c)(1)(a)]}
& \cmark
& \xmark
& \cmark
& \cmark
& \xmark
& \xmark
& \cmark
& \cmark
& \cmark
& \cmark
\\
\midrule
\makecell[l]{Disclosure of right to request deletion of personal information [CCPA §1798.105(b), 1798.130(a)(5)(A), \\ Regs \textsection{}999.308(c)(2)(a)]}
& \cmark
& \xmark
& \cmark
& \cmark
& \xmark
& \xmark
& \cmark
& \cmark
& \cmark
& \xmark
\\
\midrule
\makecell[l]{Disclosure of whether personal information is sold and right to opt-out of sale [Regs \textsection{}999.308(c)(3)(a), \\ 999.308(c)(3)(b), 999.306]}
& \cmark
& \xmark
& \cmark
& \cmark
& \xmark
& \xmark
& \cmark
& \cmark
& \cmark
& \xmark
\\
\midrule
\makecell[l]{Disclosure of right to not be discriminated against when requesting any rights [CCPA \textsection{}1798.130(a)(5)(A), \\
1798.125(a), Regs \textsection{}999.308(c)(4)(a)] }
& \xmark
& \xmark
& \cmark
& \cmark
& \xmark
& \xmark
& \cmark
& \cmark
& \cmark
& \xmark
\\
\midrule
\makecell[l]{Instructions for submitting requests and link to online form or portal if offered [Regs \textsection{}999.308(c)(1)(b), \\ 999.308(c)(2)(b), 999.308(c)(2)(c)]}
& \cmark
& \xmark
& \cmark
& \cmark
& \xmark
& \xmark
& \cmark
& \cmark
& \cmark
& \xmark
\\
\midrule
\makecell[l]{Instructions for authorized agents to make requests [Regs \textsection{}999.308(c)(5)(a)] }
& \cmark
& \xmark
& \cmark
& \cmark
& \xmark
& \xmark
& \cmark
& \cmark
& \cmark
& \xmark
\\
\midrule
\makecell[l]{Description of the process used to verify requests [Regs \textsection{}999.308(c)(1)(c)]}
& \cmark
& \xmark
& \cmark
& \cmark
& \xmark
& \xmark
& \cmark
& \cmark
& \cmark
& \xmark
\\
\midrule
\makecell[l]{List of categories of personal information collected in preceding 12 months [CCPA \textsection{}1798.130(a)(5)(B), \\1798.110(c), Regs \textsection{}999.308(c)(1)(d)]}
& \xmark
& \xmark
& \cmark
& \cmark
& \xmark
& \xmark
& \cmark
& \cmark
& \cmark
& \xmark
\\
\midrule
\makecell[l]{List of categories of personal information sold in preceding 12 months [CCPA \textsection{}1798.130(a)(5)(C), \\ 1798.115(c)(1), Regs \textsection{}999.308(c)(1)(g)(1)]}
& \cmark
& \xmark
& \cmark
& \cmark
& \xmark
& \xmark
& \cmark
& \xmark
& \cmark
& \xmark
\\
\midrule
\makecell[l]{List of categories of personal information disclosed for business purpose in preceding 12 months [CCPA \\ \textsection{}1798.130(a)(5)(C), 1798.115(c)(2), Regs \textsection{}999.308(c)(1)(g)(1)]}
& \cmark
& \xmark
& \cmark
& \cmark
& \xmark
& \xmark
& \cmark
& \xmark
& \cmark
& \xmark
\\
\midrule
\makecell[l]{For each personal information category, categories of third parties to whom information was disclosed \\or sold [Regs \textsection{}999.308(c)(1)(g)(2)]}
& \cmark
& \xmark
& \cmark
& \cmark
& \xmark
& \xmark
& \cmark
& \xmark
& \cmark
& \xmark
\\
\midrule
\makecell[l]{Categories of sources from which personal information is collected [Regs \textsection{}999.308(c)(1)(e)]}
& \cmark
& \xmark
& \cmark
& \cmark
& \xmark
& \xmark
& \cmark
& \cmark
& \cmark
& \xmark
\\
\midrule
\makecell[l]{Business or commercial purpose for collecting or selling personal information [Regs \textsection{}999.308(c)(1)(f)]}
& \cmark
& \xmark
& \cmark
& \cmark
& \xmark
& \xmark
& \cmark
& \xmark
& \cmark
& \cmark
\\
\midrule
\makecell[l]{Whether the business has actual knowledge that it sells personal information of minors under 16 years \\ of age and special process [Regs \textsection{}999.308(c)(1)(g)(3), 999.308(c)(9)]}
& \cmark
& \cmark
& \xmark
& \cmark
& \cmark
& \cmark
& \cmark
& \xmark
& \cmark
& \xmark
\\
\midrule
\makecell[l]{Contact information for questions or concerns [Regs \textsection{}999.308(c)(6)(a)]}
& \cmark
& \cmark
& \cmark
& \cmark
& \cmark
& \cmark
& \cmark
& \cmark
& \cmark
& \xmark
\\
\midrule
\makecell[l]{Date policy was last updated [Regs \textsection{}999.308(c)(7)]}
& \cmark
& \cmark
& \cmark
& \cmark
& \xmark
& \xmark
& \cmark
& \cmark
& \cmark
& \xmark
\\
\midrule
\makecell[l]{Special requirements for businesses buying, receiving, selling, or sharing personal information of \\ 10,000,000 or more consumers in a calendar year [Regs \textsection{}999.308(c)(8), 999.317(g)(1)]}
& \xmark
& \xmark
& \xmark
& \xmark
& \xmark
& \xmark
& \xmark
& \xmark
& \xmark
& \xmark
\\
\midrule
\makecell[l]{For online notices, follow generally recognized industry standards, such as the W3C Web Content \\Accessibility Guidelines, version 2.1 of June 5, 2018 [Regs \textsection{}999.308(a)(2)(d)]}
& \xmark
& \xmark
& \xmark
& \xmark
& \xmark
& \xmark
& \xmark
& \xmark
& \xmark
& \xmark
\\
\midrule
\end{tabular}
}%
\end{table*}